\documentclass[10pt,twoside]{amundi_article}

\usepackage[table]{xcolor}
\usepackage{lmodern}
\usepackage{amssymb}
\usepackage{amsfonts}
\usepackage{amsmath}
\usepackage{amsbsy}
\usepackage{fancybox}
\usepackage{fancyhdr}
\usepackage{graphicx}
\usepackage[hyphens]{url}
\usepackage{arydshln}
\usepackage{multirow}
\usepackage{pdflscape}
\usepackage{tikz}
\usepackage{comment}
\usepackage{nicefrac}
\usepackage{dsfont}
\usepackage{algorithmic}
\usepackage{algorithm}

\newlength{\figurewidth}
\newlength{\figureheight}
\def\tableskip{\vskip 10pt plus 2pt minus 2pt\relax}
\def\figureskip{\vskip 10pt plus 2pt minus 2pt\relax}

\newtheorem{remark}{Remark}
\def\limfunc#1{\mathop{\rm #1}}
\def\func#1{\mathop{\rm #1}}

\DeclareMathAlphabet\mathbfcal{OMS}{cmsy}{b}{n}
\DeclareFontFamily{OT1}{pzc}{}
\DeclareFontShape{OT1}{pzc}{m}{it}{<-> s * [1.3] pzcmi7t}{}
\DeclareMathAlphabet{\mathpzc}{OT1}{pzc}{m}{it}

\newcommand{\phI}{\vphantom{\sum\nolimits_{i=1}^{n}}}

\newcommand{\phIII}{\vphantom{\displaystyle\sum}}

\newtheorem{example}{Example}

\newcommand{\TsIII}{\hspace{3pt}}
\newcommand{\TsV}{\hspace{5pt}}
\newcommand{\TsVIII}{\hspace{8pt}}
\newcommand{\TsX}{\hspace{10pt}}

\begin{document}

\setcounter{page}{1}

\title{\textbf{\color{amundi_blue}Constrained Risk Budgeting Portfolios\\Theory, Algorithms, Applications \& Puzzles}%
\footnote{We would like to thank Thibault Bourgeron, Joan Gonzalvez, Edmond
Lezmi, Jean-Tristan Marin, Sarah Perrin, Roman Rubsamen and Jiali Xu for their
helpful comments.}}

\author{{\color{amundi_dark_blue} Jean-Charles Richard} \\
Quantitative Research \\
Eisler Capital \\
\texttt{jcharles.richard@gmail.com} \and
{\color{amundi_dark_blue} Thierry Roncalli} \\
Quantitative Research \\
Amundi Asset Management, Paris \\
\texttt{thierry.roncalli@amundi.com}}

%\author{{\color{amundi_dark_blue} Jean-Charles Richard} \\
%Quantitative Research, Eisler Capital \\
%\texttt{Jean-Charles.Richard@eislercapital.com} \and
%{\color{amundi_dark_blue} Thierry Roncalli} \\
%Quantitative Research, Amundi Asset Management\\
%\texttt{thierry.roncalli@amundi.com}}

\date{\color{amundi_dark_blue}January 2019}

\maketitle

\begin{abstract}
This article develops the theory of risk budgeting portfolios, when we would
like to impose weight constraints. It appears that the mathematical problem
is more complex than the traditional risk budgeting problem. The formulation
of the optimization program is particularly critical in order to determine the right risk
budgeting portfolio. We also show that numerical solutions can be found using
methods that are used in large-scale machine learning problems. Indeed, we
develop an algorithm that mixes the method of cyclical coordinate descent
(CCD), alternating direction method of multipliers (ADMM), proximal operators
and Dykstra's algorithm. This theoretical body is then applied to some
investment problems. In particular, we show how to dynamically control the
turnover of a risk parity portfolio and how to build smart beta portfolios
based on the ERC approach by improving the liquidity of the portfolio or
reducing the small cap bias. Finally, we highlight the importance of the
homogeneity property of risk measures and discuss the related scaling puzzle.
\end{abstract}

\medskip


\section{Introduction}

Since the 2008 Global Financial Crisis, the development of risk budgeting (RB)
techniques has marked an important milestone in portfolio management by putting
diversification at the center of portfolio construction (Qian, 2005;
Maillard \textsl{et al.}, 2010). In particular, the equal
risk contribution (ERC) portfolio has been very popular and has significantly
impacted the asset management industry. For instance, this allocation approach
is extensively implemented in risk parity funds, factor investing and
alternative risk premia (Roncalli, 2017).\smallskip

The main advantages of RB portfolios are the stability of the allocation, and
the diversification management principle, which appear more robust than the
diversification mechanism of mean-variance optimized portfolios (Bourgeron
\textsl{et al.}, 2018). This is why we don't need to add constraints in order
to regularize the solution. This advantage is also its drawback. Indeed, there
are some situations where portfolio managers have to impose constraints. For
example, they may want to impose a minimum investment weight, a
sector-neutrality or some liquidity thresholds. The goal of this paper is then
to define what does a constrained risk budgeting portfolio mean, since adding
constraints will change the risk budgets that are targeted, meaning that
ex-post risk contributions are not equal to ex-ante risk budgets.\smallskip

This paper is organized as follows. Section Two illustrates the bridge between
risk budgeting and portfolio optimization. In Section Three, we present the
right mathematical formulation of constrained risk budgeting portfolios, and
develop the corresponding numerical algorithms. In Section Four, we consider
some applications in asset allocation, in particular the management of turnover
or the consideration of liquidity. Finally, we discuss the compatibility puzzle
of the homogeneity property of coherent risk measures.

\section{The original risk budgeting portfolio}

%In what follows, we recall the original formulation of RB portfolios and how
%they can be computed.

\subsection{Definition of the risk budgeting portfolio}

We consider a universe of $n$ risky assets. Let $\mu $ and $\Sigma $ be the
vector of expected returns and the covariance matrix of asset returns. We have
$\Sigma _{i,j}=\rho _{i,j}\sigma _{i}\sigma _{j}$ where $\sigma _{i}$ is the
volatility of asset $i$ and $\rho _{i,j}$ is the correlation between asset $i$
and asset $j$. Following Roncalli (2015), we consider the standard
deviation-based risk measure defined as follows:%
\begin{equation}
\mathcal{R}\left( x\right) = -x^{\top }\left( \mu -r\right) +c\cdot \sqrt{%
x^{\top }\Sigma x} \label{eq:sdb}
\end{equation}%
where $c$ is a scalar that measures the trade-off between the expected return
of the portfolio and its volatility. We deduce that the risk
contribution of Asset $i$ is given by:%
\begin{equation*}
\mathcal{RC}_{i}\left( x\right) =x_{i}\cdot \left( -\left( \mu _{i}-r\right)
+c\frac{\left( \Sigma x\right) _{i}}{\sqrt{x^{\top }\Sigma x}}\right)
\end{equation*}%
Following Maillard \textsl{et al.} (2010), Roncalli (2013) defines the risk
budgeting (RB) portfolio using the following non-linear system:
\begin{equation}
\left\{
\begin{array}{l}
\mathcal{RC}_{i}\left( x\right) =b_{i}\mathcal{R}\left( x\right)  \\
b_{i}>0 \\
x_{i}\geq 0 \\
\sum_{i=1}^{n}b_{i}=1 \\
\sum_{i=1}^{n}x_{i}=1%
\end{array}%
\right.   \label{eq:rb1}
\end{equation}%
where $b_{i}$ is the risk budget of Asset $i$ expressed in relative terms. The
constraint $b_{i}>0$ implies that we cannot set some risk budgets to zero. This
restriction is necessary in order to ensure that the RB portfolio is unique.

\begin{remark}
Roncalli (2015) shows that the existence of the RB portfolio depends on the
value taken by $c$. In particular, the RB portfolio exists and is unique if $%
c>\func{SR}^{+}$ where $\func{SR}^{+}$ is the maximum Sharpe ratio of the
asset universe:%
\begin{equation*}
\limfunc{SR}\nolimits^{+}=\max \left( \sup_{x\in \left[ 0,1\right] ^{n}}\func{SR}\left(
x\mid r\right) ,0\right)
\end{equation*}
\end{remark}

\begin{remark}
The original ERC portfolio is obtained by considering the volatility risk
measure and the same risk budgets. It is equivalent to seting $\mu _{i}=r$, $c=1
$ and $b_{i}=1/n$. In this case, we have:%
\begin{equation*}
\mathcal{RC}_{i}\left( x\right) =\frac{x_{i}\cdot \left( \Sigma x\right) _{i}%
}{\sqrt{x^{\top }\Sigma x}}=\frac{1}{n}\sqrt{x^{\top }\Sigma x}
\end{equation*}
\end{remark}

\subsection{The associated optimization problem}

\subsubsection{The wrong formulation}

System (\ref{eq:rb1}) is equivalent to solving $n$ non-linear equations with
$n$ unknown variables. Therefore, we can use the Newton or Broyden methods
to find the numerical solution. We also deduce that:%
\begin{equation*}
\frac{1}{b_{i}}\mathcal{RC}_{i}\left( x\right) =\frac{1}{b_{j}}\mathcal{RC}%
_{j}\left( x\right) \qquad \text{for all }i,j
\end{equation*}%
In order to find the solution, an alternative approach is to solve the
optimization problem:%
\begin{eqnarray}
x_{\mathrm{RB}} & = & \arg \min \sum\limits_{i=1}^{n}\sum\limits_{j=1}^{n}\left( \frac{1}{%
b_{i}}\mathcal{RC}_{i}\left( x\right) -\frac{1}{b_{j}}\mathcal{RC}_{j}\left(
x\right) \right) ^{2} \label{eq:problem1} \\
& \text{s.t.} & \left\{
\begin{array}{l}
\mathbf{1}^{\top }x=1 \\
x\geq \mathbf{0}%
\end{array}%
\right.   \notag
\end{eqnarray}%
At the optimum $x_{\mathrm{RB}}$, the objective function $f\left(x\right)$ must
be equal to zero. This approach was originally proposed by Maillard
\textsl{et al.} (2010) in the case $b_{i}=b_{j}$.\smallskip

At first sight, Problem (\ref{eq:problem1}) seems to be easy to solve because
it resembles how a quadratic functions in $n$ variables is defined and we can
analytically compute the gradient vector and the hessian matrix of the objective
function. In fact, it is not a convex problem (Feng and Palomar, 2015), and the
optimization is tricky when the number of assets is large. From a theoretical
point of view, the objective function is not well defined because the solution
is only valid if the zero can be reached: $f\left(x_{\mathrm{RB}}\right) = 0$.
But the most important issue is that the optimization problem is driven by the
equality constraint: $\mathbf{1}^{\top }x=1$. Indeed, if we remove it, the
solution is $x_{\mathrm{RB}}=\mathbf{0}$.

\subsubsection{The right formulation}

Roncalli (2013) shows that the RB portfolio is the solution
of the following optimization problem:
\begin{eqnarray}
x_{\mathrm{RB}} &=&\arg \min \mathcal{R}\left( x\right)   \label{eq:problem2}
\\
&\text{s.t.}&\left\{
\begin{array}{l}
\sum_{i=1}^{n}b_{i}\ln x_{i}\geq \kappa ^{\star } \\
\mathbf{1}^{\top }x=1 \\
x\geq \mathbf{0}%
\end{array}%
\right.   \notag
\end{eqnarray}%
where $\kappa ^{\star }$ is a constant to be determined. This optimization
program is equivalent to finding the optimal solution $x^{\star }\left( \kappa
\right) $:%
\begin{eqnarray}
x^{\star }\left( \kappa \right)  &=&\arg \min \mathcal{R}\left( x\right)
\label{eq:problem3} \\
&\text{s.t.}&\left\{
\begin{array}{l}
\sum_{i=1}^{n}b_{i}\ln x_{i}\geq \kappa  \\
x\geq \mathbf{0}%
\end{array}%
\right.   \notag
\end{eqnarray}%
where $\kappa $ is an arbitrary constant and to scale the solution:%
\begin{equation*}
x_{\mathrm{RB}}=\frac{x^{\star }\left( \kappa \right) }{\mathbf{1}^{\top
}x^{\star }\left( \kappa \right) }
\end{equation*}%
Using the Lagrange formulation, we obtain an equivalent solution:%
\begin{eqnarray}
x^{\star }\left( \lambda \right)  &=&\arg \min \mathcal{R}\left( x\right)
-\lambda \sum_{i=1}^{n}b_{i}\ln x_{i}  \label{eq:problem4} \\
&\text{s.t.}&x\geq \mathbf{0}  \notag
\end{eqnarray}%
where $\lambda $ is an arbitrary positive scalar and:%
\begin{equation*}
x_{\mathrm{RB}}=\frac{x^{\star }\left( \lambda \right) }{\mathbf{1}^{\top
}x^{\star }\left( \lambda \right) }
\end{equation*}
$x^{\star }\left( \lambda \right) $ is the solution of a standard logarithmic
barrier problem, which has very appealing characteristics. First, it defines a
unique solution. Second, the constraint $\mathbf{1}^{\top }x=1$ is removed,
meaning that the optimization exploits the scaling property. Finally, the
constraint $x\geq \mathbf{0}$ is redundant since the logarithm is defined for
strictly positive numbers.\smallskip

We claim that Problem (\ref{eq:problem4}) is the right risk budgeting problem.
For instance, Maillard \textsl{et al.} (2010) used this formulation to show
that the ERC portfolio exists and is unique. Roncalli (2013) also noticed that
there is a discontinuity when one or more risk budgets $b_{i}$ are equal to
zero. In this case, we can find several solutions that satisfy $\mathcal{RC}%
_{i}\left( x\right) =b_{i}\mathcal{R}\left( x\right) $ or Problem
(\ref{eq:problem1}), but only one solution if we consider the logarithmic
barrier program.

\subsection{Numerical solution}

\subsubsection{The Newton algorithm}

Spinu (2013) proposes solving Problem (\ref{eq:problem4}) by using the Newton
algorithm\footnote{\label{footnote:newton1}The first and second derivatives are
computed using the following analytical expressions:
\begin{eqnarray*}
\frac{\partial \,f\left( x\right) }{\partial \,x_{i}} &=&-\left( \mu
_{i}-r\right) +c\frac{\left( \Sigma x\right) _{i}}{\sqrt{x^{\top }\Sigma x}}%
-\lambda \frac{b_{i}}{x_{i}} \\
\frac{\partial ^{2}\,f\left( x\right) }{\partial \,x_{i}\,\partial \,x_{j}}
&=&c\frac{\rho _{i,j}\sigma _{i}\sigma _{j}\sqrt{x^{\top }\Sigma x}-\left(
\Sigma x\right) _{i}\left( \Sigma x\right) _{j}}{x^{\top }\Sigma x} \\
\frac{\partial ^{2}\,f\left( x\right) }{\partial \,x_{i}^{2}} &=&c\frac{%
\sigma _{i}^{2}\sqrt{x^{\top }\Sigma x}-\left( \Sigma x\right) _{i}^{2}}{%
x^{\top }\Sigma x}+\lambda \frac{b_{i}}{x_{i}^{2}}
\end{eqnarray*}%
}:%
\begin{equation*}
x^{\left( k+1\right) }=x^{\left( k\right) }-\eta ^{\left( k\right) }\left(
\frac{\partial ^{2}\,f\left( x^{\left( k\right) }\right) }{\partial
\,x\,\partial \,x^{\top }}\right) ^{-1}\frac{\partial \,f\left( x^{\left(
k\right) }\right) }{\partial \,x}
\end{equation*}%
where $\eta ^{\left( k\right) }\in \left[ 0,1\right] $ is the step size and $k$
is the iteration index. Generally, we set $\eta ^{\left( k\right) }=1$. Spinu
(2013) noticed that the Newton algorithm may be improved because the risk
measure is self-concordant. In this case, we can use the results of Nesterov
(2004) to determine the optimal size $\eta ^{\left( k\right) }$ at each
iteration.

\subsubsection{The CCD algorithm}

The descent algorithm is defined by the following rule:%
\begin{eqnarray*}
x^{\left( k+1\right) } &=&x^{\left( k\right) }+\Delta x^{\left( k\right) } \\
&=&x^{\left( k\right) }-\eta D^{\left( k\right) }
\end{eqnarray*}%
At the $k^{\mathrm{th}}$ Iteration, the current solution $x^{\left( k\right) }$
is updated by going in the opposite direction to $D^{\left( k\right) }$. For
instance, $D^{\left( k\right) }$ is equal respectively to $\partial
_{x}\,f\left( x^{\left( k\right) }\right) $ in the gradient algorithm and $%
\left( \partial _{x,x}^{2}\,f\left( x^{\left( k\right) }\right) \right)
^{-1}\partial _{x}\,f\left( x^{\left( k\right) }\right) $ in the Newton
algorithm. Coordinate descent is a modification of the descent algorithm by
minimizing the function along one coordinate at each step:%
\begin{eqnarray*}
x_{i}^{\left( k+1\right) } &=&x_{i}^{\left( k\right) }+\Delta x_{i}^{\left(
k\right) } \\
&=&x_{i}^{\left( k\right) }-\eta D_{i}^{\left( k\right) }
\end{eqnarray*}%
The coordinate descent algorithm becomes a scalar problem, and we know that
minimizing a function with respect to one variable is easier than with $n$
variables. Concerning the choice of the variable $i$, there are two approaches:
random coordinate descent or RCD (Nesterov, 2012) and cyclical coordinate
descent or CCD (Tseng, 2001). In the first case, we assign a random number
between $1$ and $n$ to the index $i $. In the second case, we cyclically
iterate through the coordinates:
\begin{equation*}
x_{i}^{\left( k+1\right) }=\underset{x}{\arg \min }f\left( x_{1}^{\left(
k+1\right) },\ldots ,x_{i-1}^{\left( k+1\right) },x,x_{i+1}^{\left( k\right)
},\ldots ,x_{n}^{\left( k\right) }\right)
\end{equation*}%
This ensures that all the indices are selected during one cycle. In the CCD
algorithm, $k$ is the cycle index while $i$ is the iteration index within a
cycle.\smallskip

Griveau-Billion \textsl{et al.} (2013) propose applying the CCD algorithm to
find the solution of the objective function:
\begin{equation*}
f\left( x\right) =-x^{\top }\pi +c\sqrt{x^{\top }\Sigma x}-\lambda
\sum_{i=1}^{n}b_{i}\ln x_{i}
\end{equation*}%
where $\pi =\mu -r$. The first-order condition is:
\begin{equation*}
\frac{\partial \,\mathcal{L}\left( x;\lambda \right) }{\partial \,x_{i}}%
=-\pi _{i}+c\frac{\left( \Sigma x\right) _{i}}{\sigma \left( x\right) }%
-\lambda \frac{b_{i}}{x_{i}}
\end{equation*}%
At the optimum, we have $\partial _{x_{i}}\,\mathcal{L}\left( x;\lambda \right)
=0$ or:
\begin{equation*}
c\sigma _{i}^{2}x_{i}^{2}+\left( c\sigma _{i}\sum_{j\neq i}x_{j}\rho
_{i,j}\sigma _{j}-\pi _{i}\sigma \left( x\right) \right) x_{i}-\lambda
b_{i}\sigma \left( x\right) =0
\end{equation*}%
By definition of the RB portfolio we have $x_{i}>0$. We notice that the
polynomial function is convex because we have $\sigma _{i}^{2}>0$. Since the
product of the roots is negative, we always have two solutions with opposite
signs. It can be deduced that the solution is the positive root of the second-degree
equation. For the cycle $k+1$ and the $i^{\mathrm{th}}$ coordinate, we
have:
\begin{equation*}
x_{i}=\frac{-c\left( \sigma _{i}\sum_{j\neq i}x_{j}\rho _{i,j}\sigma
_{j}\right) +\pi _{i}\sigma \left( x\right) +\sqrt{\left( c\left( \sigma
_{i}\sum_{j\neq i}x_{j}\rho _{i,j}\sigma _{j}\right) -\pi _{i}\sigma \left(
x\right) \right) ^{2}+4\lambda cb_{i}\sigma _{i}^{2}\sigma \left( x\right) }%
}{2c\sigma _{i}^{2}}
\end{equation*}%
In this equation, we have the following correspondence: $x_{i}\rightarrow
x_{i}^{\left( k+1\right) }$, $x_{j}\rightarrow x_{j}^{\left( k+1\right) }$
if $j<i$, $x_{j}\rightarrow x_{j}^{\left( k\right) }$ if $j>i$, and $%
x\rightarrow \left( x_{1}^{\left( k+1\right) },\ldots ,x_{i-1}^{\left(
k+1\right) },x_{i}^{\left( k\right) },x_{i+1}^{\left( k\right) },\ldots
,x_{n}^{\left( k\right) }\right) $. If the values of $\left( x_{1},\ldots
,x_{n}\right) $ are strictly positive and if $c>\func{SR}^{+}$,
$x_{i}^{\left(k+1\right)}$ should be strictly positive. The positivity of the
solution is then achieved after each iteration and each cycle if the starting
values are positive. Therefore, the coordinate-wise descent algorithm consists
in iterating the previous equation, and we can show that it always converges
(Roncalli, 2015).

\begin{remark}
As noted by Griveau-Billion \textsl{et al.} (2013), the previous algorithm can
be simplified by setting $\lambda $ equal to $1$ and by rescaling the solution
once the convergence is obtained. Our experience shows that it is better to
rescale the solution once the CCD algorithm has converged rather than after
each cycle. Moreover, Griveau-Billion \textsl{et al.} (2013) derive analytical
formulas in order to update $\sigma\left(x\right)$ and $\sum_{j\neq i}x_{j}\rho
_{i,j}\sigma _{j}$ at each iteration.
\end{remark}

\section{Theory of constrained risk budgeting portfolio}

\subsection{Mathematical issues}

If we consider the definition of Roncalli (2013), introducing constraints leads
to the following formulation of the constrained risk budgeting
portfolio\footnote{%
We assume that $b_{i}>0$ and $\sum_{i=1}^{n}b_{i}=1$.}:%
\begin{equation*}
\left\{
\begin{array}{l}
\mathcal{RC}_{i}\left( x\right) =b_{i}\mathcal{R}\left( x\right)  \\
x\in \mathcal{S} \\
x\in \Omega
\end{array}%
\right.
\end{equation*}%
where $\mathcal{S}$ is the standard simplex:
\begin{equation*}
\mathcal{S}=\left\{ x_{i}\geq 0:\sum_{i=1}^{n}x_{i}=1\right\}
\end{equation*}%
and $x\in \Omega $ is the set of additional constraints. Let $x^{\star }\left(
\mathcal{S}\right) $ be the risk budgeting portfolio, i.e. the
solution such that $x\in \mathcal{S}$, and $x^{\star }\left( \mathcal{S}%
,\Omega \right) $ be the constrained risk budgeting portfolio, i.e. the
solution such that $x\in \mathcal{S}$ and $x\in \Omega $. Since $x^{\star
}\left( \mathcal{S}\right) $ is unique, we deduce that the solution $%
x^{\star }\left( \mathcal{S},\Omega \right) $ exists only if $x^{\star }\left(
\mathcal{S}\right) \in \Omega $, and we have $x^{\star }\left(
\mathcal{S},\Omega \right) =x^{\star }\left( \mathcal{S}\right) $. Since we
generally have $x^{\star }\left( \mathcal{S}\right) \notin \Omega $, we deduce
that there is almost certainly no solution.\smallskip

This is why professionals generally replace the equality constraint by an
approximate equality:%
\begin{equation*}
\left\{
\begin{array}{l}
\mathcal{RC}_{i}\left( x\right) \approx b_{i}\mathcal{R}\left( x\right)  \\
x\in \mathcal{S} \\
x\in \Omega
\end{array}%
\right.
\end{equation*}%
Therefore, the optimization problem becomes:%
\begin{eqnarray}
x^{\star }\left( \mathcal{S},\Omega \right)  &=&\arg \min
\sum_{i=1}^{n}\sum_{j=1}^{n}\left( \frac{1}{b_{i}}\mathcal{RC}_{i}\left(
x\right) -\frac{1}{b_{j}}\mathcal{RC}_{j}\left( x\right) \right) ^{2} \label{eq:wrong1} \\
&\text{s.t.}&x\in \mathcal{S}\cap \Omega \notag
\end{eqnarray}%
Bai \textsl{et al.} (2016) show that this optimization problem can be
simplified
as follows:%
\begin{eqnarray}
\left\{ x^{\star }\left( \mathcal{S},\Omega \right) ,\theta ^{\star
}\right\}  &=&\arg \min \sum_{i=1}^{n}\left( \frac{1}{b_{i}}\mathcal{RC}%
_{i}\left( x\right) -\theta \right) ^{2} \label{eq:wrong2} \\
&\text{s.t.}&x\in \mathcal{S}\cap \Omega \notag
\end{eqnarray}

\begin{example}
\label{ex:wrong1}
We consider a universe of four assets. Their volatilities are
equal to $10\%$, $15\%$, $20\%$ and $30\%$. The correlation matrix of asset
returns is given by the following matrix:
\begin{equation*}
\rho =\left(
\begin{array}{cccc}
1.00 &  &  &  \\
0.50 & 1.00 &  &  \\
0.50 & 0.50 & 1.00 &  \\
0.50 & 0.50 & 0.75 & 1.00%
\end{array}%
\right)
\end{equation*}
\end{example}

\begin{table}[tbph]
\centering
\caption{Computation of ERC and RB portfolios}
\label{tab:wrong1}
\tableskip
\begin{tabular}{c|cccc|cccc} \hline
\multirow{2}{*}{Asset}   & \multicolumn{4}{c|}{ERC portfolio} & \multicolumn{4}{c}{RB portfolio} \\
& $x_i$ & $\mathcal{MR}_i$ & $\mathcal{RC}_i$ & $\mathcal{RC}_i^{\star}$ &
$x_i$ & $\mathcal{MR}_i$ & $\mathcal{RC}_i$ & $\mathcal{RC}_i^{\star}$ \\ \hline
$1$ & $41.01$ & ${\TsV}7.79$ & ${\TsV}3.19$ & $25.00$ & $45.05$ & ${\TsV}8.06$ & ${\TsV}3.63$ & $30.00$ \\
$2$ & $27.34$ &      $11.68$ & ${\TsV}3.19$ & $25.00$ & $30.04$ &      $12.09$ & ${\TsV}3.63$ & $30.00$ \\
$3$ & $18.99$ &      $16.82$ & ${\TsV}3.19$ & $25.00$ & $14.67$ &      $16.10$ & ${\TsV}2.36$ & $19.50$ \\
$4$ & $12.66$ &      $25.23$ & ${\TsV}3.19$ & $25.00$ & $10.24$ &      $24.23$ & ${\TsV}2.48$ & $20.50$ \\ \hline
\multicolumn{1}{c}{$\sigma\left(x\right)$} & & & $12.78$ & \multicolumn{1}{c}{} & & & $12.11$ &         \\ \hline
\end{tabular}
\end{table}

\begin{table}[tbph]
\centering
\caption{Computation of ERC and RB portfolios when $x_i \leq 30\%$}
\label{tab:wrong2}
\tableskip
\begin{tabular}{c|cccc|cccc} \hline
\multirow{2}{*}{Asset}   & \multicolumn{4}{c|}{ERC portfolio} & \multicolumn{4}{c}{RB portfolio} \\
& $x_i$ & $\mathcal{MR}_i$ & $\mathcal{RC}_i$ & $\mathcal{RC}_i^{\star}$ &
$x_i$ & $\mathcal{MR}_i$ & $\mathcal{RC}_i$ & $\mathcal{RC}_i^{\star}$ \\ \hline
$1$ & $30.00$ & ${\TsV}7.19$ & ${\TsV}2.16$ & $15.50$ & $30.00$ & ${\TsV}7.19$ & ${\TsV}2.16$ & $15.48$ \\
$2$ & $30.00$ &      $11.60$ & ${\TsV}3.48$ & $24.98$ & $30.00$ &      $11.60$ & ${\TsV}3.48$ & $24.96$ \\
$3$ & $24.57$ &      $17.43$ & ${\TsV}4.28$ & $30.74$ & $24.43$ &      $17.42$ & ${\TsV}4.26$ & $30.52$ \\
$4$ & $15.43$ &      $25.98$ & ${\TsV}4.01$ & $28.78$ & $15.57$ &      $26.01$ & ${\TsV}4.05$ & $29.04$ \\ \hline
\multicolumn{1}{c}{$\sigma\left(x\right)$} & & & $13.93$ & \multicolumn{1}{c}{} & & & $13.94$ &         \\ \hline
\end{tabular}
\end{table}

By using the volatility risk measure, we compute the ERC portfolio and the RB
portfolio corresponding to the risk budgets $\left(30\%, 30\%, 19.5\%,
20.5\%\right)$. In Table \ref{tab:wrong1}, we report the solution of ERC and RB
portfolios. We also indicate the marginal risk $\mathcal{MR}_{i}$, the absolute
risk contribution $\mathcal{RC}_{i}$ and the relative risk contribution
$\mathcal{RC}_{i}^{\star }$. Let us now introduce the constraint $x_i \leq
30\%$ and solve the optimization problem (\ref{eq:wrong2}). Since this
constraint is not satisfied by the previous unconstrained portfolio, it has an
impact as shown in Table \ref{tab:wrong2}. As expected, the relative risk
contributions (or ex-post risk budgets) are completely different from the
ex-ante risk budgets. The concept of \textquotedblleft equal risk
contribution\textquotedblright\ does not make any sense. Moreover, we observe
that the ordering relationship between risk budgets are not preserved when we
introduce constraints. For example, in the case of the RB portfolio, we have
$b_3 < b_4$ ($19.50\%$ versus $20.50\%$) but $\mathcal{RC}_{3}^{\star
}>\mathcal{RC}_{4}^{\star }$ ($30.52\%$ versus $29.04\%$) although the
allocation in Assets 3 and 4 does not reach the upper bound constraint. We also
notice that the constrained ERC portfolio is very close to the constrained RB
portfolio. This gives us the feeling that the choice of risk budgets has little
impact, and the solution is mainly driven by the constraints.

\subsection{Formulation of the optimization problem}

Like for the unconstrained risk budgeting portfolio, we argue that the right
optimization problem is:
\begin{eqnarray}
x^{\star }\left( \mathcal{S},\Omega \right)  &=&\arg \min \mathcal{R}\left(
x\right)   \label{eq:crb1} \\
&\text{s.t.}&\left\{
\begin{array}{l}
\sum_{i=1}^{n}b_{i}\ln x_{i}\geq \kappa ^{\star } \\
x\in \mathcal{S}\cap \Omega
\end{array}%
\right.   \notag
\end{eqnarray}%
where $\kappa ^{\star }$ is a constant to be determined. We notice that the
previous problem can be simplified because:
\begin{enumerate}
\item the logarithmic barrier constraint imposes that $x_{i}\geq 0$;

\item Roncalli (2013) shows that there is only one constant $\kappa ^{\star }
    $ such that the constraint $\mathbf{1}^{\top }x=1$ is satisfied.
\end{enumerate}
We can then consider the following optimization problem:%
\begin{eqnarray}
x^{\star }\left(\Omega ,\kappa \right)  &=&\arg \min \mathcal{R}%
\left( x\right)   \label{eq:crb2} \\
&\text{s.t.}&\left\{
\begin{array}{l}
\sum_{i=1}^{n}b_{i}\ln x_{i}\geq \kappa  \\
x\in \Omega
\end{array}%
\right.   \notag
\end{eqnarray}%
and we have:%
\begin{equation*}
x^{\star }\left( \mathcal{S},\Omega \right) =\left\{ x^{\star }\left(
\Omega ,\kappa ^{\star }\right) :\sum_{i=1}^{n}x_{i}^{\star }\left(
\Omega ,\kappa ^{\star }\right) =1\right\}
\end{equation*}%
This new formulation is appealing since the constraint $x\in \mathcal{S}$ is
not explicit, but it is implicitly embedded in the optimization problem. From a
computational point of view, this reduces the complexity of the numerical
algorithm. When the constraint $x\in \Omega $ vanishes, we retrieve the
previous scaling rule. Otherwise, we consider the Lagrange formulation:
\begin{eqnarray}
x^{\star }\left(\Omega ,\lambda \right)  &=&\arg \min \mathcal{R%
}\left( x\right) -\lambda \sum_{i=1}^{n}b_{i}\ln x_{i}  \label{eq:crb3} \\
&\text{s.t.}&x\in \Omega   \notag
\end{eqnarray}%
Again, we have:%
\begin{equation*}
x^{\star }\left( \mathcal{S},\Omega \right) =\left\{ x^{\star }\left(
\Omega ,\lambda ^{\star }\right) :\sum_{i=1}^{n}x_{i}^{\star }\left(
\Omega ,\lambda ^{\star }\right) =1\right\}
\end{equation*}%
\smallskip

Formulations (\ref{eq:crb1}), (\ref{eq:crb2}) and (\ref{eq:crb3}) have the
advantage of revealing the true nature of risk budgeting. The objective is to
minimize the risk measure subject to a penalization (Richard and Roncalli,
2015). A risk budgeting portfolio is then a minimum risk portfolio subject to
hard risk budgeting, constraints, whereas a constrained risk budgeting
portfolio is a minimum risk portfolio subject to soft risk
budgeting constraints. Because of the convexity of the optimization problem
(\ref{eq:crb2}), it follows that:
\begin{equation}
\kappa _{2}\geq \kappa _{1}\Rightarrow \mathcal{R}\left( x^{\star }\left(
\Omega ,\kappa _{2}\right) \right) \geq \mathcal{R}\left(
x^{\star }\left( \Omega ,\kappa _{1}\right) \right)
\label{eq:property1}
\end{equation}%
for a given set of constraints $\Omega $. This property is fundamental since it
is the essence of risk budgeting, and it explains the relationships between
long-only minimum variance, risk budgeting and equally-weighted portfolios
obtained by Maillard \textsl{et al.} (2010) and the relationships between
long-only minimum risk, risk budgeting and weight-budgeting portfolios obtained
by Roncalli (2013). This property is also necessary to impose the continuity
of risk budgeting portfolios in particular when some risk budgets tend to zero.
Without this property, it is impossible to show that the RB portfolio is
unique, and to determine the true solution in the case where there are several
solutions to Problem (\ref{eq:rb1}) when $b_{i}=0$.

\begin{remark}
Imposing tighter constraints does not necessarily increase the risk measure%
\footnote{Indeed, we have:
\begin{equation*}
\Omega _{2}\subset \Omega _{1}\Rightarrow \mathcal{R}\left( x^{\star }\left(
\Omega _{2},\kappa \right) \right) \geq \mathcal{R}\left(
x^{\star }\left(\Omega _{1},\kappa \right) \right)
\end{equation*}%
for a given value of $\kappa $. However, the value $\kappa ^{\star }$ to
obtain the optimal portfolio $x^{\star }\left( \mathcal{S},\Omega\right)$ depends on $\Omega $.}:%
\begin{equation*}
\Omega _{2}\subset \Omega _{1}\nRightarrow \mathcal{R}\left( x^{\star
}\left( \mathcal{S},\Omega _{2}\right) \right) \geq \mathcal{R}\left(
x^{\star }\left( \mathcal{S},\Omega _{1}\right) \right)
\end{equation*}
\end{remark}

\begin{figure}[tbh]
\centering
\caption{Volatility of the constrained RB portfolio when $x_1 \leq x_1^{-}$}
\label{fig:wrong4}
\figureskip
\includegraphics[width = \figurewidth, height = \figureheight]{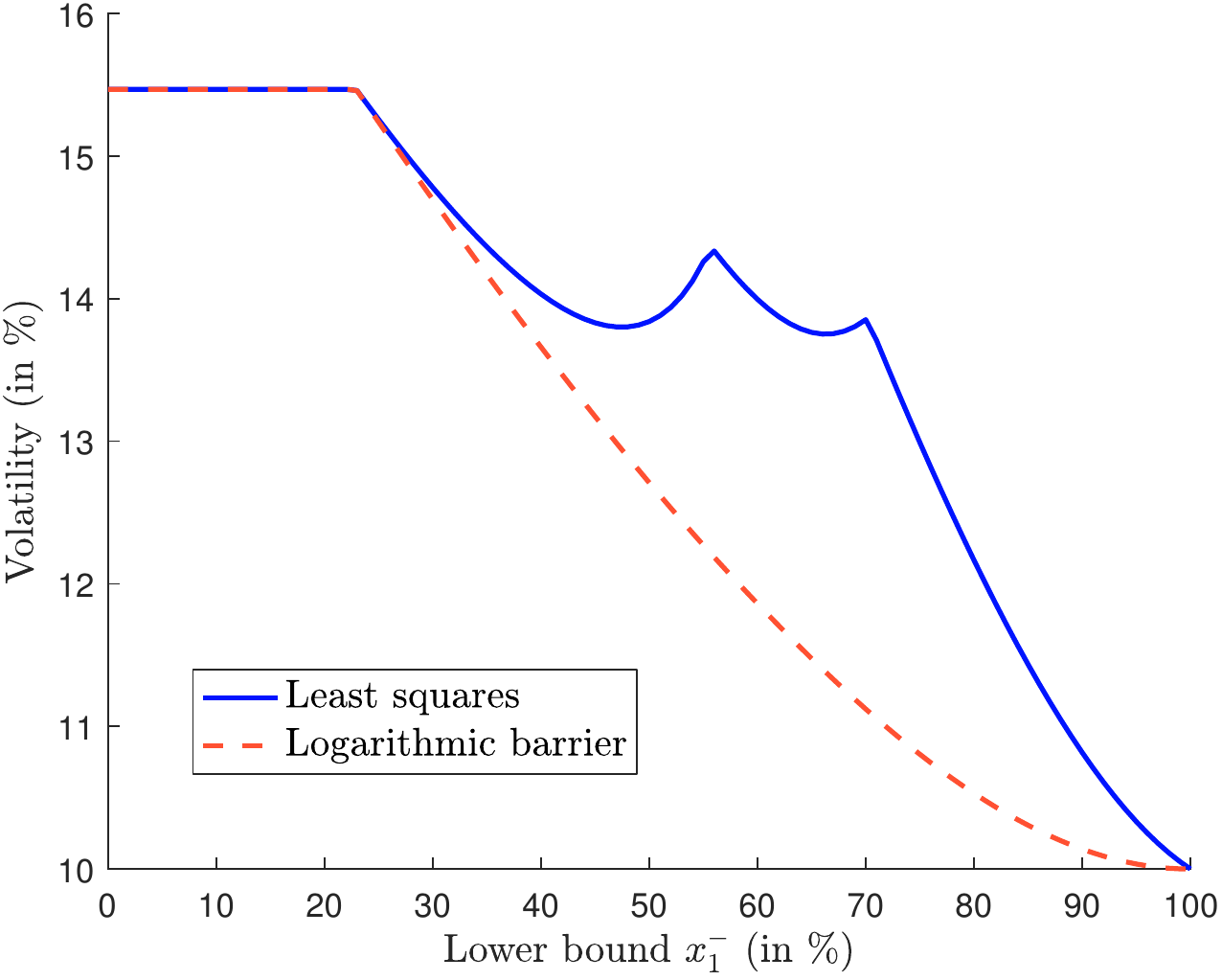}
\end{figure}

In order to illustrate the risk measure issue, we consider Example
\ref{ex:wrong1} with the risk budgets $b=\left( 10\%,20\%,30\%,40\%\right) $.
Moreover, we impose that the weight $x_{1}$ of Asset 1 is greater than a given
lower bound $x_{1}^{-}$. In Figure \ref{fig:wrong4}, we have reported the
portfolio volatility $\sigma \left( x^{\star }\left( \mathcal{S},\Omega \right)
\right) $ of the constrained risk budgeting with respect to $x_{1}^{-}$. If we
consider the least squares problem (\ref{eq:wrong2}), the volatility is
non-monotonous. This is not the case if we consider the logarithmic barrier problem
(\ref{eq:crb3}). Indeed, the least squares formulation only considers the
dimension of risk contribution matching, but not the dimension of risk measure
minimization.

\subsection{Numerical algorithms}

The optimization function becomes:%
\begin{equation}
\mathcal{L}\left( x;\lambda \right) =\mathcal{R}\left( x\right) -\lambda
\sum_{i=1}^{n}b_{i}\ln x_{i}+\mathds{1}_{\Omega }\left( x\right)
\label{eq:numeric1}
\end{equation}%
where $\mathds{1}_{\Omega }\left( x\right) $ is the convex indicator function
of $\Omega $, meaning that $\mathds{1}_{\Omega }\left( x\right) =0$ for $x\in
\Omega $ and $\mathbf{1}_{\Omega }\left( x\right) =+\infty $ for $ x\notin
\Omega $. The choice of $\lambda $ is very important, since the constrained RB
portfolio is obtained for the optimal value $\lambda ^{\star } $ such that the
sum of weights is equal to one. This can be done using the Newton-Raphson or
the bisection algorithm. If we note $x^{\star }\left(
\lambda \right) $ the solution of the minimization problem (\ref{eq:numeric1}%
), we obtain Algorithm \ref{alg:algorithm1} in the case of the bisection
method.

\begin{algorithm}[tbph]
\begin{algorithmic}
\STATE The goal is to compute the optimal Lagrange multiplier $\lambda ^{\star }$ and the solution $x^{\star }\left( \mathcal{S},\Omega \right) $
\STATE We consider two scalars $a_{\lambda }$ and $b_{\lambda }$ such that $a_{\lambda }<b_{\lambda }$ and $\lambda^{\star }\in \left[ a_{\lambda },b_{\lambda }\right] $
\STATE We note $\varepsilon _{\lambda }$ the convergence criterion of the bisection algorithm (e.g. $10^{-8}$)
\REPEAT
    \STATE We calculate $\lambda =\dfrac{a_{\lambda }+b_{\lambda }}{2}$
    \STATE We compute $x^{\star }\left( \lambda \right) $ the solution of the minimization problem:
            \begin{equation*}
                x^{\star }\left( \lambda \right) =\arg \min \mathcal{L}\left( x;\lambda\right)
            \end{equation*}
    \IF{$\sum_{i=1}^{n}x_{i}^{\star }\left( \lambda \right) <1$}
    \STATE $a_{\lambda}\leftarrow \lambda $
    \ELSE
    \STATE $b_{\lambda }\leftarrow \lambda $
    \ENDIF
\UNTIL{$\left\vert \sum\limits_{i=1}^{n}x_{i}^{\star }\left( \lambda \right) -1\right\vert \leq \varepsilon _{\lambda }$}
\RETURN $\lambda ^{\star }\leftarrow \lambda $ and $x^{\star }\left( \mathcal{S},\Omega \right) \leftarrow x^{\star }\left(\lambda^{\star} \right) $
\end{algorithmic}
\caption{General algorithm for computing the constrained RB portfolio}
\label{alg:algorithm1}
\end{algorithm}

\subsubsection{ADMM algorithm}

In order to solve Problem (\ref{eq:numeric1}), we exploit the separability
of $\mathcal{L}\left( x;\lambda \right) $. For example, we can write:%
\begin{equation}
\mathcal{L}\left( x;\lambda \right) =\underset{f\left( x\right) }{%
\underbrace{\mathcal{R}\left( x\right) -\lambda
\sum\nolimits_{i=1}^{n}b_{i}\ln x_{i}}} \quad + \quad \underset{g\left( x\right) }{%
\underbrace{\phI \quad \mathds{1}_{\Omega }\left( x\right) \quad}}  \label{eq:numeric2}
\end{equation}%
or:%
\begin{equation}
\mathcal{L}\left( x;\lambda \right) =\underset{f\left( x\right) }{%
\underbrace{\phI \mathcal{R}\left( x\right) +\mathds{1}_{\Omega }\left( x\right) }%
} \quad + \quad \underset{g\left( x\right) }{\underbrace{-\lambda
\sum\nolimits_{i=1}^{n}b_{i}\ln x_{i}}}  \label{eq:numeric3}
\end{equation}%
We notice that we have:%
\begin{eqnarray}
\left\{ x^{\star }\left( \lambda \right) ,z^{\star }\left( \lambda \right)
\right\}  &=&\arg \min f\left( x\right) +g\left( z\right)
\label{eq:numeric4} \\
&\text{s.t.}&x-z=0  \notag
\end{eqnarray}%
It follows that we can use the alternative direction method of multipliers
(ADMM) to solve this optimization problem. Algorithm \ref{alg:algorithm2}
describes the different steps.\smallskip

In the case of the Lagrange function (\ref{eq:numeric2}), the $x$-update is
equivalent to solving a penalized risk budgeting problem whereas the $z$%
-update corresponds to a proximal operator. Therefore, we can use the Newton
algorithm\footnote{\label{footnote:newton2}The first and second derivatives of
$f^{\left( k\right) }\left( x\right) =f\left( x\right) +\dfrac{\varphi
}{2}\left\Vert x-z^{\left( k-1\right) }+u^{\left( k-1\right) }\right\Vert
_{2}^{2}$ are equal to:
\begin{eqnarray*}
\frac{\partial \,f^{\left( k\right) }\left( x\right) }{\partial \,x_{i}} &=&%
\frac{\partial \,f\left( x\right) }{\partial \,x_{i}}+\varphi \left(
x_{i}-z^{\left( k-1\right) }+u^{\left( k-1\right) }\right)  \\
\frac{\partial ^{2}\,f^{\left( k\right) }\left( x\right) }{\partial
\,x_{i}\,\partial \,x_{j}} &=&\frac{\partial ^{2}\,f\left( x\right) }{%
\partial \,x_{i}\,\partial \,x_{j}} \\
\frac{\partial ^{2}\,f^{\left( k\right) }\left( x\right) }{\partial
\,x_{i}^{2}} &=&\frac{\partial ^{2}\,f\left( x\right) }{\partial \,x_{i}^{2}}%
+\varphi
\end{eqnarray*}%
where the derivatives of $f\left( x\right)$ are given in Footnote
\ref{footnote:newton1} on page \pageref{footnote:newton1}.} to find $x^{\left(
k\right) }$. For the $z$-update, we have:
\begin{equation*}
z^{\left( k\right) } =\mathbf{prox}_{g/\varphi}\left( v\right) =\arg \min\nolimits_{z}\left\{
g\left( z\right) +\frac{\varphi}{2}\left\Vert z-v_{z}^{\left( k\right) }\right\Vert _{2}^{2}\right\}
\end{equation*}%
where $v_{z}^{\left( k\right) }=x^{\left( k\right) }+u^{\left( k-1\right) }$.
If we assume that $g\left( z\right) =\mathds{1}_{\Omega }\left( z\right) $
where $\Omega $ is a convex set, we obtain:%
\begin{eqnarray*}
z^{\left( k\right) }  &=&\arg \min\nolimits_{z}\left\{ \mathds{1}%
_{\Omega }\left( z\right) +\frac{\varphi}{2}\left\Vert z-v_{z}^{\left( k\right) }\right\Vert
_{2}^{2}\right\}  \\
&=&\mathcal{P}_{\Omega }\left( v_{z}^{\left( k\right) }\right)
\end{eqnarray*}%
where $\mathcal{P}_{\Omega }\left( v\right) $ is the standard projection. In
Appendix \ref{appendix:section-proximal} on page
\pageref{appendix:section-proximal}, we give the results for the generic
constraints that we encounter in portfolio optimization. We develop some
special cases in the next section.\smallskip

\begin{algorithm}[t]
\caption{ADMM algorithm for computing the portfolio $x^{\star }\left(\lambda \right)$}
\label{alg:algorithm2}
\begin{algorithmic}
\STATE The goal is to compute the solution $x^{\star }\left(\Omega, \lambda \right) $ for a given value of $\lambda$
\STATE We initialize $x^{\left( 0\right) }$ and we choose $0\leq \varphi \leq 1$
\STATE We set $z^{\left( 0\right) }=x^{\left( 0\right) }$ and $u^{\left( 0\right) }=\mathbf{0}$
\STATE We note $\varepsilon$ the convergence criterion of the ADMM algorithm (e.g. $10^{-8}$)
\REPEAT
\STATE \begin{eqnarray*}
x^{\left( k\right) } &=&\arg \min \left\{ f\left( x\right) +\frac{\varphi }{2%
}\left\Vert x-z^{\left( k-1\right) }+u^{\left( k-1\right) }\right\Vert
_{2}^{2}\right\}  \\
z^{\left( k\right) } &=&\arg \min \left\{ g\left( z\right) +\frac{\varphi }{2%
}\left\Vert x^{\left( k\right) }-z+u^{\left( k-1\right) }\right\Vert
_{2}^{2}\right\}  \\
u^{\left( k\right) } &=&u^{\left( k-1\right) }+\left( x^{\left( k\right)
}-z^{\left( k\right) }\right)
\end{eqnarray*}%
\UNTIL{$\left\Vert x^{\left( k\right) }-z^{\left( k\right) }\right\Vert \leq \varepsilon $}
\RETURN $x^{\star }\left( \lambda \right) \leftarrow x^{\left(k\right) }$
\end{algorithmic}
\end{algorithm}

For the Lagrange function (\ref{eq:numeric3}), the $x$-update
corresponds to a constrained risk minimization problem whereas the $z$-update
is equivalent to solving a penalized logarithmic barrier problem. The $x$-update
can be done using constrained non-linear optimization methods\footnote{In order
to accelerate the convergence, we can implement analytical derivatives, which
are the same than those given in Footnote \ref{footnote:newton2} on page
\pageref{footnote:newton2} by setting $\lambda =0$ in the derivatives of the
function $f\left(x\right)$.}, whereas the $z$-step corresponds to the proximal
operator of
the logarithmic barrier function\footnote{%
See Appendix \ref{appendix:section-log-barrier} on page \pageref{appendix:section-log-barrier}.}:%
\begin{equation*}
z_{i}^{\left( k\right) }=\frac{\varphi \left( x_{i}^{\left( k\right)
}+u_{i}^{\left( k-1\right) }\right) +\sqrt{\varphi ^{2}\left( x_{i}^{\left(
k\right) }+u_{i}^{\left( k-1\right) }\right) ^{2}+4\varphi \lambda b_i}}{%
2\varphi }
\end{equation*}%
If we consider the volatility risk measure $\mathcal{R}\left( x\right) =%
\sqrt{x^{\top }\Sigma x}$ instead of the standard deviation-based risk measure given by Equation (\ref{eq:sdb}),
the ADMM algorithm is simplified as follows\footnote{We have:%
\begin{eqnarray*}
f^{\left( k\right) }\left( x\right)  &=&\frac{1}{2}x^{\top }\Sigma x+\frac{%
\varphi }{2}\left( x-z^{\left( k-1\right) }+u^{\left( k-1\right) }\right)
^{\top }\left( x-z^{\left( k-1\right) }+u^{\left( k-1\right) }\right)  \\
&=&\frac{1}{2}x^{\top }\Sigma x+\frac{\varphi }{2}\left( x^{\top }x-2x^{\top
}v_{x}^{\left( k\right) } + \left( v_{x}^{\left( k\right) }\right) ^{\top }v_{x}^{\left( k\right) } \right)
\end{eqnarray*}%
where $v_{x}^{\left( k\right) } =z^{\left( k-1\right) }-u^{\left( k-1\right) }$.}:%
\begin{eqnarray*}
x^{\left( k\right) } &=&\arg \min \frac{1}{2}x^{\top }\left( \Sigma +\varphi
I_{n}\right) x-\varphi x^{\top }\left( z^{\left( k-1\right) }-u^{\left(
k-1\right) }\right)  \\
&\text{s.t.}&x\in \Omega  \\
z^{\left( k\right) } &=&\frac{\varphi \left( x_{i}^{\left( k\right)
}+u_{i}^{\left( k-1\right) }\right) +\sqrt{\varphi ^{2}\left( x_{i}^{\left(
k\right) }+u_{i}^{\left( k-1\right) }\right) ^{2}+4\varphi \lambda b}}{%
2\varphi } \\
u^{\left( k\right) } &=&u^{\left( k-1\right) }+\left( x^{\left( k\right)
}-z^{\left( k\right) }\right)
\end{eqnarray*}
This algorithm exploits the property that minimizing the portfolio volatility
is equivalent to minimizing the portfolio variance, even if this last risk measure
does not satisfy the Euler decomposition. If $\Omega$ is a set of linear
(equality and inequality) constraints, the $x$-update reduces to a standard QP
problem.

\subsubsection{CCD algorithm}

Another route for solving Problem (\ref{eq:numeric2}) is to consider the CCD
algorithm. Convergence of coordinate descent methods requires that the function
is strictly convex and differentiable. However, Tseng (2001) has extended the
convergence properties to a non-differentiable class of functions:
\begin{equation*}
f\left( x\right) =f_{0}\left( x\right) +\sum_{i=1}^{n}f_{i}\left( x_i\right)
\end{equation*}%
where $f_{0}$ is strictly convex and differentiable and the functions $f_{i}$
are non-differentiable. Dealing with convex constraints is equivalent to
writing the constraints in the sum term of $f\left( x\right) $. If we consider
the formulation (\ref{eq:numeric2}) and the standard deviation-based risk
measure, we have:
\begin{equation*}
\mathcal{L}\left( x;\lambda \right) =\mathcal{L}_{0}\left( x;\lambda \right)
+\mathbf{1}_{\Omega }\left( x\right)
\end{equation*}%
where $\mathcal{L}_{0}\left( x;\lambda \right)$ is defined as follows:
\begin{equation}
\mathcal{L}_{0}\left( x;\lambda \right) =-x^{\top }\pi +c\sqrt{x^{\top
}\Sigma x}-\lambda \sum_{i=1}^{n}b_{i}\ln x_{i}  \notag
\end{equation}%

\paragraph{The case of separable constraints}
If we assume that the set of constraints is separable with respect to all the
variables:
\begin{equation*}
\Omega =\bigcap_{i=1}^{n}\Omega _{i}
\end{equation*}%
where $\Omega _{i}$ is the constraint on $x_{i}$, we have $f_{i}\left( x_{i}\right) =\mathds{1}_{\Omega _{i}}\left(
x_{i}\right) $. We deduce that the CCD
algorithm consists in two steps. We first solve the minimization problem of
$\mathcal{L}_{0}\left( x;\lambda \right) $ for one coordinate, and then we
compute the projection onto $\Omega _{i}$. For the first step, the first-order condition is:%
\begin{equation*}
\frac{\partial \,\mathcal{L}_{0}\left( x;\lambda \right) }{\partial \,x_{i}}%
=-\pi _{i}+c\frac{\left( \Sigma x\right) _{i}}{\sqrt{x^{\top }\Sigma x}}%
-\lambda \frac{b_{i}}{x_{i}}=0
\end{equation*}%
It follows that $cx_{i}\left( \Sigma x\right) _{i}-\pi _{i}x_{i}\sigma \left( x\right)
-\lambda b_{i}\sigma \left( x\right) =0$ or equivalently:
\begin{equation*}
\alpha _{i}x_{i}^{2}+\beta _{i}x_{i}+\gamma _{i}=0
\end{equation*}%
where:%
\begin{equation*}
\left\{
\begin{array}{l}
\alpha _{i}=c\sigma _{i}^{2} \\
\beta _{i}=c\sigma _{i}\sum_{j\neq i}x_{j}\rho _{i,j}\sigma _{j}-\pi
_{i}\sigma \left( x\right)  \\
\gamma _{i}=-\lambda b_{i}\sigma \left( x\right)
\end{array}%
\right.
\end{equation*}%
We deduce that the coordinate solution is the positive root of the
second-degree equation:
\begin{equation*}
x_{i} = \frac{-\beta _{i}+\sqrt{\beta _{i}^{2}-4\alpha _{i}\gamma _{i}%
}}{2\alpha _{i}}
\end{equation*}%
The second step is the projection into the set $\Omega _{i}$:%
\begin{equation*}
x_{i} = \mathcal{P}_{\Omega _{i}}\left( x_{i}\mathcal{}\right)
\end{equation*}%
Finally, we obtain Algorithm \ref{alg:algorithm3}.

\begin{algorithm}[tbh]
\caption{CCD algorithm for computing the portfolio $x^{\star }\left(\lambda
\right)$ when the set of constraints is separable with respect to the variables $x_i$}
\label{alg:algorithm3}
\begin{algorithmic}
\STATE The goal is to compute the solution $x^{\star }\left(\Omega, \lambda \right) $ for a given value of $\lambda $
\STATE We initialize the vector $x$
\STATE We note $\varepsilon$ the convergence criterion of the CCD algorithm (e.g. $10^{-8}$)
\REPEAT
    \STATE $x^{\prime} \leftarrow x$
    \FOR {$i=1:n$}
        \STATE $\sigma_x \leftarrow \sigma\left(x\right)$
        \STATE  We update $x_{i}$ as follows:
                \begin{equation*}
                    x_{i} \leftarrow \frac{-\beta_{i}+\sqrt{\beta _{i}^{2}-4\alpha _{i} \gamma _{i}}}{2\alpha_{i}}
                \end{equation*}
                where:
                \begin{equation*}
                    \left\{
                    \begin{array}{l}
                        \alpha _{i} = c\sigma _{i}^{2} \\
                        \beta _{i}  = c\sigma _{i}\sum_{j\neq i}x_{j}\rho _{i,j}\sigma _{j}-\pi _{i}\sigma_x  \\
                        \gamma _{i} = -\lambda b_{i}\sigma_x
                    \end{array}
                    \right.
                \end{equation*}
        \STATE $x_i \leftarrow \mathcal{P}_{\Omega_i }\left( x_i\right)$
    \ENDFOR
\UNTIL{$\sum_{i=1}^{n}\left( x_{i}^{\prime }-x_{i}\right) ^{2}\leq \varepsilon$}
\RETURN $x^{\star }\left( \lambda \right) \leftarrow x^{\left( k\right) }$
\end{algorithmic}
\end{algorithm}

\begin{remark}
Our algorithm differs from the one given by Nesterov (2012) and Wright (2015). Let $\eta >0$ be
the stepsize of the gradient descent. The coordinate update is:
\begin{equation*}
x_{i}^{\star }=\arg \min \left( x-x_{i}\right) g_{i}+\frac{1}{2\eta }\left(
x-x_{i}\right) ^{2}+\xi \cdot \mathds{1}_{\Omega _{i}}\left( x\right)
\end{equation*}%
where $\xi $ is a positive scalar and:%
\begin{equation*}
g_{i}=-\pi _{i}+c\frac{\left( \Sigma x\right) _{i}}{\sqrt{x^{\top }\Sigma x}}%
-\lambda \frac{b_{i}}{x_{i}}
\end{equation*}%
In our case, this algorithm is very simple\footnote{%
See Appendix \ref{appendix:section-wright} on page
\pageref{appendix:section-wright}.}, because it reduces to calculate the
proximal of $x_{i}-\eta g_{i}$ associated with the function $\mathds{1}_{\Omega
_{i}}\left( x\right)$. However, we prefer to use the previous algorithm in
order to exploit the analyticity of the solution.
\end{remark}

\paragraph{The case of non-separable constraints}

A first idea is to replace the projection step $x_{i}\leftarrow
\mathcal{P}_{\Omega _{i}}\left( x_{i}\right) $ by something equivalent that
ensures that the constraints are verified. The natural approach is to apply the
proximal operator or equivalently the projection: $x\leftarrow
\mathcal{P}_{\Omega }\left( x\right) $. In practice, we observe that the CCD
solution does not always converge to the true solution. It will depend on how
the constraints and the variables are ordered. In fact, the true approach is to
use a block-coordinate algorithm if constraints are separable by blocks. This
is not always the case. This is why we prefer to use the previous ADMM
algorithms or the ADMM-CCD algorithm, which is described in the next paragraph.

\begin{algorithm}[tbph]
\caption{ADMM-CCD algorithm for computing the portfolio $x^{\star }\left(\lambda \right)$}
\label{alg:algorithm4}
\begin{algorithmic}
\STATE The goal is to compute the solution $x^{\star }\left( \Omega ,\lambda \right) $ for a given value of $\lambda $
\STATE We initialize $x^{\left( 0\right) }$ and we choose $0\leq \varphi \leq 1$
\STATE We set $z^{\left( 0\right) }=x^{\left( 0\right) }$ and $u^{\left( 0\right) }=\mathbf{0}$
\STATE We note $\varepsilon $ and $\varepsilon ^{\prime }$ the convergence criterion of ADMM and CCD algorithms
\STATE We note $k_{\max}$ the maximum number of ADMM iterations

\FOR {$k = 1 : k_{\max}$}
    \STATE \centerline{\underline{\textbf{I. $x$-update}}}
    \STATE $v_{x}^{\left( k\right) } \leftarrow z^{\left( k-1\right) }-u^{\left(k-1\right) }$
    \STATE $\tilde{x}\leftarrow x^{\left( k-1\right) }$
    \REPEAT
        \STATE $\tilde{x}^{\prime }\leftarrow \tilde{x}$
        \FOR {$i=1:n$}
            \STATE We update the volatility $\sigma _{x}\leftarrow \sigma \left( \tilde{x}\right) $ and calculate:
                \begin{equation*}
                    \left\{\begin{array}{l}
                    \alpha _{i}=c\sigma _{i}^{2}+\varphi \sigma _{x} \\
                    \beta _{i}=c\sigma _{i}\sum_{j\neq i}\tilde{x}_{j}\rho _{i,j}\sigma_{j}-\left( \pi _{i}+\varphi v_{x_{i}}^{\left( k\right) }\right) \sigma _{x} \\
                    \gamma _{i}=-\lambda b_{i}\sigma _{x}%
                \end{array}\right.
                \end{equation*}
            \STATE We update $\tilde{x}_{i}$ as follows:
                \begin{equation*}
                    \tilde{x}_{i} \leftarrow \frac{-\beta _{i}+\sqrt{\beta _{i}^{2}-4\alpha_{i}\gamma _{i}}}{2\alpha _{i}}
                \end{equation*}
        \ENDFOR
    \UNTIL{$\sum_{i=1}^{n}\left( \tilde{x}_{i}^{\prime }-\tilde{x}_{i}\right) ^{2}\leq \varepsilon ^{\prime }$}
    \STATE $x^{\left( k\right) }\leftarrow \tilde{x}$
    \STATE
    \STATE \centerline{\underline{\textbf{II. $z$-update}}}
    \STATE $v_{z}^{\left( k\right) } \leftarrow x^{\left( k\right) }+u^{\left(k-1\right) }$
    \STATE $z^{\left( k\right) } \leftarrow \mathcal{P}_{\Omega }\left( v_{z}^{\left( k\right) }\right)$
    \STATE
    \STATE \centerline{\underline{\textbf{III. $u$-update}}}
    \STATE $u^{\left( k\right) } \leftarrow  u^{\left( k-1\right) }+x^{\left(k\right) }-z^{\left( k\right) }$
    \STATE
    \STATE \centerline{\underline{\textbf{IV. Convergence test}}}
    \IF{$\left\Vert x^{\left( k\right) }-z^{\left( k\right) }\right\Vert\leq \varepsilon $}
       \STATE Break
    \ENDIF
\ENDFOR
\RETURN $x^{\star }\left( \lambda \right) \leftarrow x^{\left(k\right) }$
\end{algorithmic}
\end{algorithm}

\subsubsection{Mixed ADMM-CCD algorithm}

If we consider the formulation (\ref{eq:numeric2}), we have already shown
that the $x$-update of the ADMM algorithm corresponds to a regularized risk
budgeting problem. Therefore, we can use the CCD algorithm to find the
solution $x^{\left( k\right) }$. We remind that:%
\begin{equation*}
f^{\left( k\right) }\left( x\right) =-x^{\top }\pi +c\sqrt{x^{\top }\Sigma x}%
-\lambda \sum_{i=1}^{n}b_{i}\ln x_{i}+\dfrac{\varphi }{2}\left\Vert
x-v_{x}^{\left( k\right) }\right\Vert _{2}^{2}
\end{equation*}%
where $v_{x}^{\left( k\right) }=z^{\left( k-1\right) }-u^{\left( k-1\right) }
$. The first-order condition is:%
\begin{equation*}
\frac{\partial \,f^{\left( k\right) }\left( x\right) }{\partial \,x_{i}}%
=-\pi _{i}+c\frac{\left( \Sigma x\right) _{i}}{\sqrt{x^{\top }\Sigma x}}%
-\lambda \frac{b_{i}}{x_{i}}+\varphi \left( x_{i}-v_{x_{i}}^{\left( k\right)
}\right) =0
\end{equation*}%
It follows that:%
\begin{equation*}
cx_{i}\left( \Sigma x\right) _{i}-\pi _{i}x_{i}\sigma \left( x\right)
-\lambda b_{i}\sigma \left( x\right) +\varphi x_{i}^{2}\sigma \left(
x\right) -\varphi x_{i}v_{x_{i}}^{\left( k\right) }\sigma \left( x\right) =0
\end{equation*}%
or:%
\begin{equation*}
\alpha _{i}x_{i}^{2}+\beta _{i}x_{i}+\gamma _{i}=0
\end{equation*}%
where:%
\begin{equation*}
\left\{
\begin{array}{l}
\alpha _{i}=c\sigma _{i}^{2}+\varphi \sigma \left( x\right)  \\
\beta _{i}=c\sigma _{i}\sum_{j\neq i}x_{j}\rho _{i,j}\sigma _{j}-\left( \pi
_{i}+\varphi v_{x_{i}}^{\left( k\right) }\right) \sigma \left( x\right)  \\
\gamma _{i}=-\lambda b_{i}\sigma \left( x\right)
\end{array}%
\right.
\end{equation*}%
We deduce that the coordinate solution is the positive root of the previous
second-degree equation:
\begin{equation*}
x_{i}=\frac{-\beta _{i}+\sqrt{\beta _{i}^{2}-4\alpha _{i}\gamma _{i}%
}}{2\alpha _{i}}
\end{equation*}%
Finally, we obtain the ADMM-CCD algorithm, which is described on the previous
page.

\subsubsection{Efficiency of the algorithms}

We may investigate the efficiency of the previous algorithms. Firstly, our
experience shows that traditional constrained optimization algorithms (SQP,
trust region, constrained interior point, etc.) fail to find the solution. It
is somewhat surprising because we have the feeling that the optimization
problem of constrained risk budgeting portfolios seems to be standard.
Unfortunately, this is not the case, because the mixing of the logarithmic
barrier and constraints is not usual. This is why it is important to implement
the previous algorithms. Secondly, all the algorithms are not equal and the
implementation is key in particular when we consider large problems with more
than one hundred assets. In Table \ref{tab:box2}, we have reported the
computational time we have obtained for solving the example described on page
\pageref{section:dynamic-allocation}. For that, we consider five different
methods: ADMM-Newton, ADMM-BFGS, ADMM-QP, ADMM-CCD and CCD. For each method, we
consider three implementations:

\begin{enumerate}
\item The first one considers that the primal variable $\varphi $ is constant
    ($\varphi = 1$) and we use the classical bisection method described in Algorithm \ref{alg:algorithm1}.

\item The second one considers that the penalization variable $\varphi $ is
    constant ($\varphi = 1$), and we use an accelerated bisection method. The
    underlying idea is to choose a starting value $x^{\left(0\right)}$ for
    the ADMM/CCD algorithm, which is not constant and depends on the Lagrange
    coefficient $\lambda$. The corresponding method is described in Appendix
    \ref{appendix:section-bisection} on page
    \pageref{appendix:section-bisection}.

\item The third uses the accelerated bisection algorithm, and considers the
    adaptive method for the variable $\varphi ^{\left( k\right) }$, which is
    given in Appendix \ref{appendix:section-adaptive} on page
    \pageref{appendix:section-adaptive}. The underlying idea is to accelerate
    the convergence of the ADMM algorithm by using the right scale of the
    primal residual variable $u^{\left(k\right)}$.

\end{enumerate}
Results using our Matlab implementation are reported in Table \ref{tab:box2}.
Absolute figures are not interesting, because they depend on the processing
power of the computer. Relative figures show that computational times can
differ dramatically from one algorithm to another, from one implementation to
another. The best algorithms are ADMM-Newton and CCD, followed by ADMM-CCD.
Curiously, the ADMM-QP method is less efficient%
\footnote{This is due to the \texttt{quadprog} procedure of Matlab. Indeed, we
do not observe same results in Python.}. Finally, the worst algorithm is the
ADMM-BFGS algorithm. However, we notice a large improvement in this algorithm
if we implement the accelerated bisection and the adaptive method for scaling
the regularization parameter. Indeed, the computational time is divided by a
factor of 15!

\begin{table}[tbph]
\centering
\caption{Computational time using our Matlab implementation (relative value)}
\label{tab:box2}
\tableskip
\begin{tabular}{ccccc} \hline
Algorithm & $x$-update &         (1) &         (2) &   (3) \\ \hline
ADMM      & Newton     &  ${\TsX}2$ &   ${\TsX}1$ &  ${\TsX}1$ \\
ADMM      & BFGS       &      $380$ &       $280$ & ${\TsV}25$ \\
ADMM      & QP         &      $220$ &       $120$ &      $110$ \\
ADMM      & CCD        & ${\TsV}10$ &   ${\TsX}9$ &  ${\TsX}8$ \\
CCD       &            &  ${\TsX}1$ &   ${\TsX}1$ &            \\ \hline
\end{tabular}
\end{table}

\subsection{Special cases}

\subsubsection{Box constrained optimization}
\label{section:box}

In portfolio optimization, imposing lower and upper bounds is frequent:
\begin{equation*}
\Omega =\left\{ x \in \mathbb{R}^n : x^{-}\leq x\leq x^{+}\right\}
\end{equation*}
For example, box constraints are used when limiting single exposures because of regulatory constraints or controlling
the turnover of the portfolio.

\paragraph{The optimization framework}

In the box constrained case, the Lagrange function becomes:%
\begin{eqnarray}
\mathcal{L}\left( x;\lambda ,\lambda ^{-},\lambda ^{+}\right)  &=&-x^{\top
}\pi +c\sqrt{x^{\top }\Sigma x}-\lambda \sum_{i=1}^{n}b_{i}\ln x_{i}-  \notag
\\
&&\sum_{i=1}^{n}\lambda _{i}^{-}\left( x_{i}-x_{i}^{-}\right)
-\sum_{i=1}^{n}\lambda _{i}^{+}\left( x_{i}^{+}-x_{i}\right)
\label{eq:box1}
\end{eqnarray}%
The first-order condition is:%
\begin{equation*}
\frac{\partial \,\mathcal{L}\left( x;\lambda ,\lambda ^{-},\lambda
^{+}\right) }{\partial \,x_{i}}=-\pi _{i}+c\frac{\left( \Sigma x\right) _{i}%
}{\sqrt{x^{\top }\Sigma x}}-\lambda \frac{b_{i}}{x_{i}}-\lambda
_{i}^{-}+\lambda _{i}^{+}=0
\end{equation*}%
We deduce that:%
\begin{equation*}
\mathcal{RC}_{i}\left( x\right) =\lambda b_{i}+\lambda _{i}^{-}x_{i}-\lambda
_{i}^{+}x_{i}
\end{equation*}%
Since the Kuhn-Tucker conditions are:%
\begin{equation*}
\left\{
\begin{array}{l}
\min \left( \lambda _{i}^{-},x_{i}-x_{i}^{-}\right) =0 \\
\min \left( \lambda _{i}^{+},x_{i}^{+}-x_{i}\right) =0%
\end{array}%
\right.
\end{equation*}%
we obtain three cases:
\begin{enumerate}
\item If no bound is reached, we have $\lambda _{i}^{-}=0$ and $\lambda
    _{i}^{+}=0$, and we retrieve the RB portfolio $x_{\mathrm{RB}}$;
\item If the lower bound is reached, $\lambda _{i}^{-}>0$ and the risk
    contribution of Asset $i$ is higher than $b_{i}$;
\item If the upper bound is reached, $\lambda _{i}^{+}>0$ and the risk
    contribution of Asset $i$ is lower than $b_{i}$.
\end{enumerate}

\begin{remark}
When we compare the logarithmic barrier solution with the least squares
solution of the ERC portfolio, we observe that the first one preserves the
\textquotedblleft equal risk contribution\textquotedblright\ property for all
the assets that do not reach lower or upper bounds. This is not the case with
the least squares solution, since the risk contribution is different for all
the assets.
\end{remark}

The previous algorithms require the proximal operator to be computed:%
\begin{eqnarray*}
x^{\star } &=&\mathbf{prox}_{g}\left( \tilde{x}\right)  \\
&=&\mathcal{P}_{\Omega }\left( \tilde{x}\right)
\end{eqnarray*}%
where $\tilde{x}= x^{\left( k\right) }+u^{\left( k-1\right) }$ is the
value of $v_z^{k}$ in the ADMM procedure. In Appendix \ref{appendix:section-proximal} on page
\pageref{appendix:section-proximal}, we show that the proximal operator
corresponds to the truncation operator:
\begin{equation*}
\mathbf{prox}_{g}\left( \tilde{x}\right) =\mathcal{T}\left( \tilde{x}; x^{-}, x^{+}\right)
\end{equation*}%
where:%
\begin{equation*}
\mathcal{T}\left( \tilde{x}; x^{-}, x^{+}\right) =\left\{
\begin{array}{ll}
x_{i}^{-} & \text{if }\tilde{x}_{i}<x_{i}^{-} \\
\tilde{x}_{i} & \text{if }x_{i}^{-}\leq \tilde{x}_{i}\leq x_{i}^{+} \\
x_{i}^{+} & \text{if }\tilde{x}_{i}>x_{i}^{+}%
\end{array}%
\right.
\end{equation*}%
For the CCD algorithm, the projection $x_i \leftarrow \mathcal{P}_{\Omega_i
}\left( x_i\right)$ reduces to apply the truncation operator to the single
coordinate $x_i$.\smallskip

At the optimum, we deduce that:%
\begin{equation*}
\lambda _{i}^{-\star }=\max \left( \frac{\mathcal{RC}_{i}\left( x^{\star }\left(
\mathcal{S},\Omega \right) \right) -\mathcal{\lambda }^{\star }b_{i}}{%
x_{i}^{\star }\left( \mathcal{S},\Omega \right) },0\right)
\end{equation*}%
and:%
\begin{equation*}
\lambda _{i}^{+\star }=\max \left( \frac{\mathcal{\lambda }^{\star }b_i-\mathcal{RC}%
_{i}\left( x^{\star }\left( \mathcal{S},\Omega \right) \right) }{%
x_{i}^{\star }\left( \mathcal{S},\Omega \right) },0\right)
\end{equation*}%
where $\lambda ^{\star }$ is the solution of the bisection algorithm.

\begin{remark}
In order to find the optimal value $\lambda ^{\star }$, we need an initial
guess $\lambda _{0}$ for the Newton-Raphson algorithm or an interval $\left[
a_{\lambda },b_{\lambda }\right] $ for the bisection method. Let $x_{\mathrm{%
RB}}$ be the RB portfolio without constraints. The optimal Lagrange
coefficient associated with $x_{\mathrm{RB}}$ is equal to\footnote{%
Indeed, the first-order condition is:%
\begin{eqnarray*}
\frac{\partial \,\mathcal{R}\left( x\right) }{\partial \,x_{i}}-\lambda
\frac{b_{i}}{x_{i}}=0 &\Leftrightarrow &x_{i}\frac{\partial \,\mathcal{R}%
\left( x\right) }{\partial \,x_{i}}-\lambda b_{i}=0 \\
&\Leftrightarrow &\sum_{i=1}^{n}x_{i}\frac{\partial \,\mathcal{R}\left(
x\right) }{\partial \,x_{i}}-\lambda \sum_{i=1}^{n}b_{i}=0 \\
&\Leftrightarrow &\lambda =\sum_{i=1}^{n}x_{i}\frac{\partial \,\mathcal{R}%
\left( x\right) }{\partial \,x_{i}}=\mathcal{R}\left( x\right)
\end{eqnarray*}%
because $\sum_{i=1}^{n}b_{i}=1$.}:%
\begin{equation*}
\lambda _{\mathrm{RB}} = \sum_{i=1}^{n}\mathcal{RC}_{i}\left( x_{\mathrm{RB}}\right)  = \mathcal{R}\left( x_{\mathrm{RB}}\right)
\end{equation*}%
It follows that a good initial guess is $\lambda _{0}=\mathcal{R}\left( x_{%
\mathrm{RB}}\right) $. We also notice that:%
\begin{eqnarray*}
\lambda ^{\star } &=&\sum_{i=1}^{n}\mathcal{RC}_{i}\left( x^{\star }\left(
S,\Omega \right) \right) +\sum_{i=1}^{n}\left( \lambda _{i}^{+\star
}-\lambda _{i}^{-\star }\right) x_{i}^{\star }\left( S,\Omega \right)  \\
&=&\mathcal{R}\left( x^{\star }\left( S,\Omega \right) \right)
+\sum_{i=1}^{n}\left( \lambda _{i}^{+\star }-\lambda _{i}^{-\star }\right)
x_{i}^{\star }\left( S,\Omega \right)
\end{eqnarray*}
We deduce that $a_{\lambda }=m_{a} \cdot \mathcal{R}\left(
x_{\mathrm{RB}}\right)$ and $b_{\lambda }=m_{b} \cdot \mathcal{R}\left(
x_{\mathrm{RB}}\right)$ where the parameters $m_{a}$ and $m_{b}$ depends on the
tightness of constraints\footnote{We have $m_{a}\leq 1$ and $m_{b}\geq 1$.}.
Generally, we have $\mathcal{R}\left( x^{\star }\left( S,\Omega \right) \right)
\approx \mathcal{R}\left( x_{\mathrm{RB}}\right) $, implying that the values
$m_{a}=0.5$ and $m_{b}=2.0$ are sufficient.
\end{remark}

\paragraph{An example of dynamic allocation}
\label{section:dynamic-allocation}

We consider a universe of five assets. Their volatilities are equal to $15\%$,
$20\%$, $25\%$, $30\%$ and $10\%$. The correlation matrix of asset returns is given
by the following matrix:
\begin{equation*}
\rho =\left(
\begin{array}{ccccc}
1.00 &  &  &  &  \\
0.10 & 1.00 &  &  &  \\
0.40 & 0.70 & 1.00 &  &  \\
0.50 & 0.40 & 0.80 & 1.00 &  \\
0.50 & 0.40 & 0.05 & 0.10 & 1.00%
\end{array}%
\right)
\end{equation*}%
Let us assume that the current portfolio is $x_{0}=\left(
25\%,25\%,10\%,15\%,30\%\right)$. In Table \ref{tab:box1a}, we report the
volatility breakdown of this portfolio. We notice that there are some large
differences in terms of risk contributions. In particular, the second asset has
a volatility contribution of $31.1\%$. We would like to obtain a more balanced
portfolio. Table \ref{tab:box1b} shows the results of the ERC portfolio.

\begin{table}[tbph]
\centering
\caption{Volatility breakdown (in \%) of the current portfolio}
\label{tab:box1a}
\tableskip
\begin{tabular}{|c|cccc|} \hline
Asset & $x_i$ & $\mathcal{MR}_i$ & $\mathcal{RC}_i$ & $\mathcal{RC}_i^{\star}$ \\ \hline
1 & $25.00$ &      $10.00$ & ${\TsV}2.50$ & $20.21$ \\
2 & $25.00$ &      $15.40$ & ${\TsV}3.85$ & $31.10$ \\
3 & $10.00$ &      $20.30$ & ${\TsV}2.03$ & $16.41$ \\
4 & $10.00$ &      $22.24$ & ${\TsV}2.22$ & $17.98$ \\
5 & $30.00$ & ${\TsV}5.90$ & ${\TsV}1.77$ & $14.30$ \\    \hline
\multicolumn{3}{|l}{$\sigma\left(x\right)$} & $12.37$ &    \\ \hline
\end{tabular}
\end{table}
\begin{table}[tbph]
\centering
\caption{Volatility breakdown (in \%) of the ERC portfolio}
\label{tab:box1b}
\tableskip
\begin{tabular}{|c|cccc|} \hline
Asset & $x_i$ & $\mathcal{MR}_i$ & $\mathcal{RC}_i$ & $\mathcal{RC}_i^{\star}$ \\ \hline
1 & $22.40$ &      $10.61$ & ${\TsV}2.38$ & $20.00$ \\
2 & $16.51$ &      $14.39$ & ${\TsV}2.38$ & $20.00$ \\
3 & $12.03$ &      $19.74$ & ${\TsV}2.38$ & $20.00$ \\
4 & $10.51$ &      $22.60$ & ${\TsV}2.38$ & $20.00$ \\
5 & $38.54$ & ${\TsV}6.16$ & ${\TsV}2.38$ & $20.00$ \\    \hline
\multicolumn{3}{|l}{$\sigma\left(x\right)$} & $11.88$ &    \\ \hline
\end{tabular}
\end{table}

We notice that the ERC portfolio is relatively far from the current portfolio. In
particular, the turnover is equal to $22.18\%$. In order to obtain a solution
closer to the current allocation, we impose that the weights cannot deviate from
the current ones by $5\%$:
\begin{equation*}
x_{0}-5\%\leq x\leq x_{0}+5\%
\end{equation*}
The underlying idea is to move from the initial portfolio to a risk budgeting
portfolio, which presents the risk parity property as much as possible. In this
case, we obtain the results in Table \ref{tab:box1c}. We notice that three
assets (\#1, \#3 and \#4) present the same contributions ($2.35\%$) because
they do not reach the bounds. On the contrary, the second and fifth assets have
respectively a higher and lower risk contribution ($2.98\%$ and $2.10\%$)
because the lower and upper bounds are reached. This solution helps to reduce
the turnover, since it is equal to $14.22\%$.

\begin{table}[tbph]
\centering
\caption{Volatility breakdown (in \%) of the constrained RB portfolio}
\label{tab:box1c}
\tableskip
\begin{tabular}{|c|cccc|cc|} \hline
Asset & $x_i$ & $\mathcal{MR}_i$ & $\mathcal{RC}_i$ & $\mathcal{RC}_i^{\star}$ &
$\lambda_{i}^{-}$ & $\lambda_{i}^{+}$ \\ \hline
1 & $22.89$ &      $10.28$ & ${\TsV}2.35$ & $19.39$ & $0.00$ & $0.00$ \\
2 & $20.00$ &      $14.90$ & ${\TsV}2.98$ & $24.55$ & $3.13$ & $0.00$ \\
3 & $11.69$ &      $20.13$ & ${\TsV}2.35$ & $19.39$ & $0.00$ & $0.00$ \\
4 & $10.42$ &      $22.57$ & ${\TsV}2.35$ & $19.39$ & $0.00$ & $0.00$ \\
5 & $35.00$ & ${\TsV}6.00$ & ${\TsV}2.10$ & $17.29$ & $0.00$ & $0.73$ \\    \hline
\multicolumn{3}{|l}{$\sigma\left(x\right)$} & $12.14$ & & \multicolumn{2}{c|}{$\lambda = 11.76$}   \\ \hline
\end{tabular}
\end{table}

A naive solution to obtain a risk parity portfolio that matches the constraints
would be to identify the assets that reach the lower and upper bounds and to
allocate the remaining weight between the other assets by imposing the same
risk contribution. In this example, the second and fifth assets do not satisfy
the constraints. Therefore, we have to allocate $45\%$ of the allocation
between the first, third and fourth assets. This naive solution is given in
Table \ref{tab:box1d}. The ERC portfolio between the three unconstrained assets
is equal to $\left(22.84\%,12.34\%,9.83\%\right)$. In this case, the risk
contributions are the same and are equal to $2.65\%$. However, the equal risk
contribution property does not hold if we consider the full portfolio, when we
take into account the constrained assets. Indeed, we obtain
$\mathcal{RC}_1=2.34\%$, $\mathcal{RC}_3=2.49\%$ and $\mathcal{RC}_4=2.21\%$.
The reason is that risk budgeting portfolios are sensitive to the asset
universe definition (Roncalli and Weisang, 2016). This is why this two-step
naive procedure does not give the right answer. In Table \ref{tab:box1d}, we
also report the least squares solution corresponding to the optimization
problem (\ref{eq:wrong2}). Again, no assets verify the equal risk contribution
property.

\begin{table}[tbph]
\centering
\caption{Volatility breakdown (in \%) of naive and least squares solutions}
\label{tab:box1d}
\tableskip
\begin{tabular}{|c|cccc|cccc|} \hline
      & \multicolumn{4}{c|}{Naive solution} & \multicolumn{4}{c|}{Least squares solution} \\
Asset & $x_i$ & $\mathcal{MR}_i$ & $\mathcal{RC}_i$ & $\mathcal{RC}_i^{\star}$ &
        $x_i$ & $\mathcal{MR}_i$ & $\mathcal{RC}_i$ & $\mathcal{RC}_i^{\star}$ \\ \hline
1 &      $22.84$ &      $10.25$ & ${\TsV}2.34$ & $19.30$  & $23.13$ &      $10.32$ & ${\TsV}2.39$ & $19.70$ \\
2 &      $20.00$ &      $14.98$ & ${\TsV}3.00$ & $24.70$  & $20.00$ &      $14.86$ & ${\TsV}2.97$ & $24.53$ \\
3 &      $12.34$ &      $20.18$ & ${\TsV}2.49$ & $20.53$  & $11.39$ &      $20.07$ & ${\TsV}2.29$ & $18.87$ \\
4 & ${\TsV}9.83$ &      $22.46$ & ${\TsV}2.21$ & $18.20$  & $10.48$ &      $22.55$ & ${\TsV}2.36$ & $19.51$ \\
5 &      $35.00$ & ${\TsV}5.99$ & ${\TsV}2.10$ & $17.28$  & $35.00$ & ${\TsV}6.02$ & ${\TsV}2.11$ & $17.39$ \\ \hline
\multicolumn{3}{|l}{$\sigma\left(x\right)$} & $12.13$ &   &         &              &      $12.11$ &         \\ \hline
\end{tabular}
\end{table}

\newpage

\begin{remark}
The previous example shows how to take into account a current allocation when
building a risk budgeting portfolio. This approach is particularly relevant
when considering dynamic rebalancing and tactical asset allocation. One of the
main advantages of the RB approach is that it produces a stable allocation.
However, it does not enable us to consider investment constraints such as the
current allocation. Introducing weight constraints allows the fund manager to
better control the portfolio construction.
\end{remark}

\subsubsection{Risk budgeting with linear constraints}

We now consider general linear constraints:%
\begin{equation*}
\Omega =\left\{ x\in \mathbb{R}^{n}:Ax=B,Cx\leq D,x^{-}\leq x\leq
x^{+}\right\}
\end{equation*}%
These constraints generalize the case of lower and upper bounds. For example,
it is common to add some (lower and upper) limits in terms of exposures by
asset classes, sectors, ratings, etc. These types of limits are implemented thanks
to inequality constraints $Cx\leq D$. Equality constraints $Ax=B$ are less
common in portfolio optimization. In Appendix \ref{appendix:section-proximal}
on page \pageref{appendix:section-proximal}, we provide some closed-form
formulas to calculate $\mathcal{P}_{\Omega }\left( x\right) $, when $\Omega $
corresponds to $Ax=B$ or $c^{\top }x\leq d $ or $x^{-}\leq x\leq x^{+}$. First,
we notice that the analytical formula only exists for the half-space constraint
$c^{\top }x\leq d$, but not for multiple inequality constraints $\Omega
=\left\{ x \in \mathbb{R}^n : Cx\leq D\right\} $ when $ \limfunc{card}\Omega
=m>1$. The underlying idea is then to break down $\Omega $ as the intersection
of $m$ half-space sets:
\begin{equation*}
\Omega =\Omega _{1}\cap \Omega _{2}\cap \cdots \cap \Omega _{m}
\end{equation*}%
where $\Omega _{j}=\left\{ x \in \mathbb{R}^n : c_{\left( j\right) }^{\top }x\leq d_{\left(
j\right) }\right\} $, $c_{\left( j\right) }^{\top }$ corresponds to the $j^{%
\mathrm{th}}$ row of $C$ and $d_{\left( j\right) }$ is the $j^{\mathrm{th}}$
element of $D$. In this case, we can apply the Dykstra's algorithm for
computing the proximal operator of $\mathds{1}_{\Omega }\left( x\right) $. This
algorithm is given on page \pageref{alg:dykstra1}. Second, mixing the
constraints is not straightforward. Again, we can break down $\Omega $ as the
intersection of three basic convex sets:
\begin{equation*}
\Omega =\Omega _{1}\cap \Omega _{2}\cap \Omega _{3}
\end{equation*}%
where $\Omega _{1}=\left\{  \in \mathbb{R}^n : Ax=B\right\} $, $\Omega
_{2}=\left\{ x \in \mathbb{R}^n : Cx\leq D\right\} $ and $\Omega _{3}=\left\{ x
\in \mathbb{R}^n : x^{-}\leq x\leq x^{+}\right\} $. Since we know how to
project each basic set, we calculate the projection $\mathcal{P}_{\Omega
}\left( x\right) $ with the Dykstra's algorithm, which is described on page
\pageref{alg:dykstra2}.

\begin{table}[tbph]
\caption{Volatility and correlation matrix of asset returns (in \%)}
\centering
\label{tab:linear1-1}
\tableskip
\begin{tabular}{cc|cccc:cccc}
\hline
\multirow{2}{*}{$\sigma_i$}
           & & 1     & 2           & 3        & 4        & 5        & 6        & 7        & 8   \\
           & & ${\TsV}5.0$ & ${\TsV}5.0$ & ${\TsV}7.0$ & $10.0$ & $15.0$ & $15.0$ & $15.0$ & $18.0$             \\ \hline
\multirow{8}{*}{$\rho_{i,j}$}
 & 1 & ${\TsIII}100$ &               &            &            &            &            &            &       \\
 & 2 & ${\TsVIII}80$ & ${\TsIII}100$ &            &            &            &            &            &       \\
 & 3 & ${\TsVIII}60$ & ${\TsVIII}40$ & $100$      &            &            &            &            &       \\
 & 4 & $-20$         & $-20$         & ${\TsV}50$ & $100$      &            &            &            &       \\ \cdashline{2-10}
 & 5 & $-10$         & $-20$         & ${\TsV}30$ & ${\TsV}60$ & $100$      &            &            &       \\
 & 6 & $-20$         & $-10$         & ${\TsV}20$ & ${\TsV}60$ & ${\TsV}90$ & $100$      &            &       \\
 & 7 & $-20$         & $-20$         & ${\TsV}20$ & ${\TsV}50$ & ${\TsV}70$ & ${\TsV}60$ & $100$      &       \\
 & 8 & $-20$         & $-20$         & ${\TsV}30$ & ${\TsV}60$ & ${\TsV}70$ & ${\TsV}70$ & ${\TsV}70$ & $100$ \\ \hline
\end{tabular}
\end{table}

\newpage

Let us consider an example of multi-asset allocation\footnote{This example is
taken from Roncalli (2013) on page 287.}. We consider a universe of eight asset
classes: (1) US 10Y Bonds, (2) Euro 10Y Bonds, (3) Investment Grade Bonds, (4)
High Yield Bonds, (5) US Equities, (6) Euro Equities, (7) Japan Equities and
(8) EM Equities. In Table \ref{tab:linear1-1}, we indicate the statistics used
to compute the optimal allocation.\smallskip

Using these figures, we calculate the risk parity portfolio, which corresponds
to the second column in Table \ref{tab:linear1-2}. We notice that the bond
allocation is equal to $76.72\%$ whereas the equity allocation is equal to
$23.28\%$. In order to increase the equity allocation, we impose that the
weight of the last four assets is greater than $30\%$. The solution is given in
the fourth column. Finally, we overweight the allocation in European assets
with respect to American assets by $5\%$ (sixth column).

\begin{table}[tbph]
\centering
\caption{The case of inequality constraints}
\label{tab:linear1-2}
\tableskip
\begin{tabular}{|c|cc:cc:cc|}
\hline
$x_{5} + x_{6} + x_{7} + x_{8}\geq 30\%$ & & & \multicolumn{2}{c:}{$\checkmark $} & \multicolumn{2}{c|}{$\checkmark $} \\
$x_{2} + x_{6} \geq x_{1} + x_{5} + 5\%$ & & & & & \multicolumn{2}{c|}{$\checkmark $}                                  \\ \hline
Asset  & $x_i$ & $\mathcal{RC}_i^{\star}$ & $x_i$ & $\mathcal{RC}_i^{\star}$ & $x_i$ & $\mathcal{RC}_i^{\star}$        \\
1 &      $26.83$ &       $12.50$ &      $25.78$ & ${\TsV}8.64$ &      $24.52$ & ${\TsV}8.16$  \\
2 &      $28.68$ &       $12.50$ &      $27.41$ & ${\TsV}8.64$ &      $28.69$ & ${\TsV}9.13$  \\
3 &      $11.41$ &       $12.50$ & ${\TsV}9.51$ & ${\TsV}8.64$ & ${\TsV}9.52$ & ${\TsV}8.61$  \\
4 & ${\TsV}9.80$ &       $12.50$ & ${\TsV}7.29$ & ${\TsV}8.64$ & ${\TsV}7.27$ & ${\TsV}8.61$  \\
5 & ${\TsV}5.61$ &       $12.50$ & ${\TsV}7.06$ &      $15.91$ & ${\TsV}6.97$ &      $15.69$  \\
6 & ${\TsV}5.90$ &       $12.50$ & ${\TsV}7.71$ &      $16.58$ & ${\TsV}7.80$ &      $16.82$  \\
7 & ${\TsV}6.66$ &       $12.50$ & ${\TsV}9.23$ &      $18.14$ & ${\TsV}9.23$ &      $18.16$  \\
8 & ${\TsV}5.11$ &       $12.50$ & ${\TsV}6.00$ &      $14.82$ & ${\TsV}6.00$ &      $14.81$  \\ \hline
\multicolumn{1}{|c|}{$\sigma\left(x\right)$ (in \%)} &
\multicolumn{2}{c:}{${\TsV}4.78$} & \multicolumn{2}{c:}{${\TsV}5.20$} & \multicolumn{2}{c|}{${\TsV}5.19$} \\ \hline
\end{tabular}
\end{table}

\begin{remark}
Algorithm \ref{alg:algorithm3} is no longer valid when the coordinates are
coupled via constraints. This is generally the case when we impose $Ax=B$ and
$Cx\leq D$. This is why we use the ADMM-Newton or ADMM-CCD algorithms for
solving this type of constrained risk budgeting problem.
\end{remark}

\section{Applications}

We consider two applications that are based on our professional experience. The
first application is the design of risk-based equity indices. Building an ERC
portfolio on the Eurostoxx 50 universe is straightforward. This is not the case
if we consider the universe of the Eurostoxx index. Indeed, this universe
contains many small cap stocks, and having the same risk contribution for small
cap and large cap stocks may induce some liquidity issues. In particular, this
type of problem occurs when considering a large universe of non-homogenous
stocks. The second application is the control of rebalancing effects. This
issue happens when we consider short-term covariance matrices. In this case,
the allocation can be very reactive, implying large turnovers. For example,
this type of situation is observed when we implement multi-asset risk parity
strategies with daily or weekly rebalancing and empirical covariance matrices
that are estimated with less than one year of historical data. These two
applications are illustrated below.

\subsection{Risk-based indexation and smart beta portfolios}

We now consider a capitalization-weighted index composed of seven stocks. The
weights are equal to $34\%$, $25\%$, $20\%$, $15\%$, $3\%$, $2\%$ and $1\%$. We
assume that the volatilities of these stocks are equal to $15\%$, $16\%$,
$17\%$, $18\%$, $19\%$, $20\%$ and $21\%$, whereas the correlation matrix of
stock returns is given by:
\begin{equation*}
\rho =\left(
\begin{array}{ccccccc}
1.00 &      &      &      &      &      &       \\
0.75 & 1.00 &      &      &      &      &       \\
0.73 & 0.75 & 1.00 &      &      &      &       \\
0.70 & 0.70 & 0.75 & 1.00 &      &      &       \\
0.65 & 0.68 & 0.69 & 0.75 & 1.00 &      &       \\
0.62 & 0.65 & 0.63 & 0.67 & 0.70 & 1.00 &       \\
0.60 & 0.60 & 0.65 & 0.68 & 0.75 & 0.80 & 1.00
\end{array}%
\right)
\end{equation*}%
As shown by Demey \textsl{et al.} (2010), the ERC portfolio defined by Maillard
\textsl{et al} (2010) is a good candidate for building a risk-based equity
index. However, an ERC index does not take into account liquidity constraints.
For instance, we notice that the ERC allocation does not take into account the
size of stocks in Table \ref{tab:smart1}. We may assume that the three last
assets are small cap stocks. In this case, it can be more realistic to
distinguish small cap and large cap stocks. A first idea is to keep the CW
weights on the small cap universe and to apply the ERC on the large cap
universe. This solution called LC-ERC (for Large Cap ERC) is presented in Table
\ref{tab:smart1}. We face an issue here, because the ERC portfolio on large cap
stocks does not depend on the full correlation matrix. Therefore, this solution
assumes that the two universes of stocks are not correlated. We have the same
issue if we consider the least squares solution (LS-ERC). A better approach is
to find the ERC portfolio by imposing that the weights of small cap stocks
are exactly equal to the corresponding CW weights. Let $\Omega_{\mathcal{SC}}$
be the universe of small cap stocks. We
have $x_{i}=x_{\mathrm{cw},i}$ if $i\in \Omega_{\mathcal{SC}}$.
This is equivalent to imposing the following weight constraints:%
\begin{equation*}
\left\{
\begin{array}{ll}
0\leq x_{i} & \text{if }i\notin \Omega_{\mathcal{SC}} \\
x_{\mathrm{cw},i}\leq x_{i}\leq x_{\mathrm{cw},i} & \text{if }i\in \Omega_{\mathcal{SC}}%
\end{array}%
\right.
\end{equation*}
The result corresponds to the C-ERC portfolio. Again, we notice that the property
of equal risk contribution is satisfied at the global level for large cap
stocks, contrary to LC-ERC and LS-ERC portfolios.

\begin{table}[tbph]
\centering
\caption{Volatility breakdown (in \%) of constrained ERC portfolios}
\label{tab:smart1}
\tableskip
\begin{tabular}{|c|cc:cc:cc:cc:cc|}
\hline
\multirow{2}{*}{Asset}
      & \multicolumn{2}{c:}{CW} & \multicolumn{2}{c:}{ERC} & \multicolumn{2}{c:}{LC-ERC} & \multicolumn{2}{c:}{LS-ERC} & \multicolumn{2}{c|}{C-ERC} \\
      & $x_i$ & $\mathcal{RC}_i^{\star}$ & $x_i$ & $\mathcal{RC}_i^{\star}$ & $x_i$ & $\mathcal{RC}_i^{\star}$
      & $x_i$ & $\mathcal{RC}_i^{\star}$ & $x_i$ & $\mathcal{RC}_i^{\star}$ \\ \hline
1 &      $34.00$ &      $32.08$ & $17.22$ &                  $14.29$ &      $25.81$ &      $23.39$ &      $26.62$ &      $24.23$ &      $25.87$ &      $23.46$ \\
2 &      $25.00$ &      $24.82$ & $15.90$ &                  $14.29$ &      $24.06$ &      $23.44$ &      $24.20$ &      $23.63$ &      $24.07$ &      $23.46$ \\
3 &      $20.00$ &      $20.92$ & $14.78$ &                  $14.29$ &      $22.44$ &      $23.44$ &      $22.09$ &      $23.08$ &      $22.46$ &      $23.46$ \\
4 &      $15.00$ &      $16.01$ & $13.83$ &                  $14.29$ &      $21.69$ &      $23.57$ &      $21.09$ &      $22.89$ &      $21.59$ &      $23.46$ \\
5 & ${\TsV}3.00$ & ${\TsV}3.10$ & $13.17$ &                  $14.29$ & ${\TsV}3.00$ & ${\TsV}3.10$ & ${\TsV}3.00$ & ${\TsV}3.10$ & ${\TsV}3.00$ & ${\TsV}3.10$ \\
6 & ${\TsV}2.00$ & ${\TsV}2.03$ & $12.86$ &                  $14.29$ & ${\TsV}2.00$ & ${\TsV}2.02$ & ${\TsV}2.00$ & ${\TsV}2.02$ & ${\TsV}2.00$ & ${\TsV}2.02$ \\
7 & ${\TsV}1.00$ & ${\TsV}1.05$ & $12.23$ &                  $14.29$ & ${\TsV}1.00$ & ${\TsV}1.05$ & ${\TsV}1.00$ & ${\TsV}1.05$ & ${\TsV}1.00$ & ${\TsV}1.05$ \\ \hline
\multicolumn{1}{|c}{$\sigma\left(x\right)$} & \multicolumn{2}{c:}{$14.50$} & \multicolumn{2}{c:}{$15.23$} &
\multicolumn{2}{c:}{$14.68$} & \multicolumn{2}{c:}{$14.66$} & \multicolumn{2}{c|}{$14.68$} \\ \hline
\end{tabular}
\end{table}

\subsection{Managing the portfolio turnover}

On page \pageref{section:dynamic-allocation}, we have considered a dynamic allocation and imposed some
constraints in order to control the turnover of the portfolio. For that, we
have used lower and upper bounds in order to impose a maximum deviation
between the current allocation and the new allocation. This is a way of controlling
the turnover. However, we can directly impose a turnover control. Let $x_{0}$
be the current allocation. The two-way turnover of Portfolio $x$ with respect
to Portfolio $x_{0}$ is defined by:
\begin{eqnarray*}
\tau\left(  x\mid x_{0}\right)    & = &\sum_{i=1}^{n}\left\vert x_{i}%
-x_{0,i}\right\vert \\
& = & \left\Vert x-x_{0}\right\Vert _{1}%
\end{eqnarray*}
It corresponds to the $\boldsymbol{\ell}_{1}$-norm of $x$ with respect to the
centroid vector $x_{0}$. Therefore, the corresponding Lagrange function is:%
\begin{equation*}
\mathcal{L}\left(  x;\lambda\right)  =\mathcal{R}\left(  x\right)
-\lambda\sum_{i=1}^{n}b_{i}\ln x_{i}+\mathds{1}_{\Omega}\left(  x\right)
\end{equation*}
where $\Omega=\left\{  x\in R:\tau\left(  x\mid x_{0}\right)  \leq\tau^{\star
}\right\}  $ and $\tau^{\star}$ is the turnover limit. If we use the previous
algorithms, the only difficulty is calculating the proximal operator%
\footnote{We use the properties of proximal operators described in Appendix
\ref{appendix:section-proximal} on page \pageref{appendix:section-proximal}.}
of $g\left(  x\right)  =\mathds{1}_{\Omega}\left(  x\right)  $:
\begin{equation*}
\mathbf{prox}_{g}\left(  x\right)  =\mathbf{prox}_{f}\left(  x-x_{0}\right)
+x_{0}%
\end{equation*}
where $f\left(  x\right)  =\mathds{1}_{\Omega^{\prime}}\left(  x\right)  $ and
$\Omega^{\prime}=\left\{  x\in R:\left\Vert x\right\Vert _{1}\leq\tau^{\star
}\right\}  $. Finally, we deduce that:%
\begin{equation*}
\mathbf{prox}_{g}\left(  x\right) =  x-\mathbf{prox}_{\tau^{\star}\max}\left(  \left\vert x-x_{0}\right\vert
\right)  \odot\func{sign}\left(  x-x_{0}\right)
\end{equation*}
where $\mathbf{prox}_{\lambda \max}\left(v\right)$ is the proximal operator
given by Equation (\ref{eq:proximal-max}) on page
\pageref{eq:proximal-max}.\smallskip

We consider the example of multi-asset allocation on page
\pageref{tab:linear1-1}. Let us assume that the current allocation is a 50/50
asset mix policy, where the weight of each asset class is $12.5\%$. In Table
\ref{tab:turnover1}, we have reported the solution for different turnover
limits $\tau^{\star}$. If the turnover limit is very low, the optimized RB
portfolio is close to the current allocation. We verify that the optimized
portfolio tends to the ERC portfolio when we increase the turnover limit.

\begin{table}[tbph]
\centering
\caption{Constrained RB portfolios (in \%) with turnover control}
\label{tab:turnover1}
\tableskip
\begin{tabular}{|c|cccccccc|}
\hline
\multirow{2}{*}{Asset}
  & \multicolumn{8}{c|}{$\tau^\star$} \\
  &  ${\TsV}0.00$ &      $10.00$ &      $20.00$ &      $30.00$ &      $40.00$ &      $50.00$ &      $60.00$ &      $70.00$ \\ \hline
1 &       $12.50$ &      $14.86$ &      $17.28$ &      $19.68$ &      $22.01$ &      $24.28$ &      $26.58$ &      $26.83$ \\
2 &       $12.50$ &      $15.14$ &      $17.72$ &      $20.32$ &      $22.99$ &      $25.72$ &      $28.42$ &      $28.68$ \\
3 &       $12.50$ &      $12.50$ &      $12.50$ &      $12.50$ &      $12.50$ &      $12.50$ &      $11.65$ &      $11.41$ \\
4 &       $12.50$ &      $12.50$ &      $12.50$ &      $12.50$ &      $12.50$ &      $11.50$ & ${\TsV}9.90$ & ${\TsV}9.80$ \\
5 &       $12.50$ &      $11.20$ & ${\TsV}9.70$ & ${\TsV}8.49$ & ${\TsV}7.27$ & ${\TsV}6.28$ & ${\TsV}5.66$ & ${\TsV}5.61$ \\
6 &       $12.50$ &      $12.02$ &      $10.36$ & ${\TsV}9.02$ & ${\TsV}7.69$ & ${\TsV}6.63$ & ${\TsV}5.95$ & ${\TsV}5.90$ \\
7 &       $12.50$ &      $12.50$ &      $11.72$ &      $10.16$ & ${\TsV}8.66$ & ${\TsV}7.47$ & ${\TsV}6.71$ & ${\TsV}6.66$ \\
8 &       $12.50$ & ${\TsV}9.28$ & ${\TsV}8.22$ & ${\TsV}7.33$ & ${\TsV}6.39$ & ${\TsV}5.62$ & ${\TsV}5.14$ & ${\TsV}5.11$ \\ \hline
$\tau\left(x^{\star} \mid x_0\right)$
  &  ${\TsV}0.00$ &      $10.00$ &      $20.00$ &      $30.00$ &      $40.00$ &      $50.00$ &      $60.00$ &      $61.02$ \\ \hline
\end{tabular}
\end{table}

\section{Discussion and limitations of constrained risk budgeting portfolios}

In this section, we discuss the concept of constrained risk budgeting
allocation, and shows that it is not natural and has some limitations.

\subsection{Coherent risk measures and the homogeneity property}

Many people believe that the risk budgeting allocation is only related to the
Euler decomposition:
\begin{equation*}
\mathcal{R}\left( x\right) =\sum_{i=1}^{n}x_{i}\frac{\partial \,\mathcal{R}%
\left( x\right) }{\partial \,x_{i}}
\end{equation*}%
implying that the risk measure is \textquotedblleft convex\textquotedblright.
However, this concept is not always well-defined, and is often confused with
the concept of coherent risk measure. Following Artzner \textsl{et al.} (1999),
a risk measure $\mathcal{R}\left( x\right) $ is said to be coherent if it
satisfies the following properties:
\begin{enumerate}
\item Subadditivity%
\begin{equation*}
\mathcal{R}\left( x_{1}+x_{2}\right) \leq \mathcal{R}\left( x_{1}\right) +%
\mathcal{R}\left( x_{2}\right)
\end{equation*}%
The risk of two portfolios should be less than adding the risk of the two
separate portfolios.

\item Homogeneity
\begin{equation*}
\mathcal{R}\left( \lambda x\right) =\lambda \mathcal{R}\left( x\right)\quad
\text{if }\lambda \geq 0
\end{equation*}%
Leveraging or deleveraging of the portfolio increases or decreases the risk
measure in the same magnitude.

\item Monotonicity%
\begin{equation*}
\text{if }x_{1}\prec x_{2}\text{, then }\mathcal{R}\left( x_{1}\right) \geq
\mathcal{R}\left( x_{2}\right)
\end{equation*}
If Portfolio $x_{2}$ has a better return than Portfolio $x_{1}$ under all
scenarios, risk measure $\mathcal{R}\left( x_{1}\right) $ should be higher
than risk measure $\mathcal{R}\left( x_{2}\right) $.

\item Translation invariance%
\begin{equation*}
\text{if }m\in \mathbb{R}\text{, then }\mathcal{R}\left( x+m\right) =%
\mathcal{R}\left( x\right) -m
\end{equation*}%
Adding a cash position of amount $m$ to the portfolio reduces the risk by
$m$.
\end{enumerate}
F\"ollmer and Schied (2002) propose replacing the homogeneity and subadditivity
conditions by a weaker condition called the convexity property:
\begin{equation*}
\mathcal{R}\left( \lambda x_{1}+\left( 1-\lambda \right) x_{2}\right) \leq
\lambda \mathcal{R}\left( x_{1}\right) +\left( 1-\lambda \right) \mathcal{R}%
\left( x_{2}\right)
\end{equation*}%
This condition means that diversification should not increase the risk. Saying
that the risk measure is convex is ambiguous. For some people, this means that
$\mathcal{R}\left( x\right) $ satisfies the convexity property of
F\"ollmer and Schied (2002), while for other people, this means that $\mathcal{%
R}\left( x\right) $ satisfies the Euler decomposition. It is true that these
two concepts are related (Tasche, 2008), but they recover two different things
(Kalkbrener, 2005). First, there is a confusion between the Euler decomposition
and the Euler allocation principle (Tasche, 2008). Second, the Euler allocation
principle does make sense only if the risk measure is subadditive (Kalkbrener,
2005). But another property is very important. Roncalli (2015) shows that risk
budgeting is valid only if the homogeneity property is satisfied ---
$\mathcal{R}\left( \lambda x\right) =\lambda \mathcal{R}\left( x\right) $ ---
because this property ensures that there is a solution and the solution is
unique. By definition, this property is related to the scaling property of the
RB portfolio when there are no constraints. When we impose some constraints, it
is obvious that the homogeneity property is valid only if the constraints
$\Omega $ are compatible with the scaling property. This is not generally true
except for some special cases. However, we don't need the homogeneity property
to be satisfied for all values $\lambda \geq 0$. We need the homogeneity
property to be met for a range of $\lambda $ around the unconstrained risk
budgeting portfolio. This is why imposing tight constraints can lead to a
numerical solution without knowing if it corresponds to the true constrained
risk budgeting portfolio.

\subsection{The scaling puzzle}

We consider the example given on page \pageref{tab:linear1-1}. If we consider
the optimized portfolio subject to the constraint $\sum_{i=5}^{8}x_{i}\geq
30\%$, we may think that it is equivalent to the optimized portfolio subject to
the constraint $\sum_{i=1}^{4}x_{i}\leq 70\%$. Results are given in Table
\ref{tab:linear1-3}, when we use the equally-weighted portfolio as the starting
value $x^{\left(0\right)}$ in the ADMM algorithms. It is surprising to obtain
two different solutions. Nevertheless, they are very close. If we do the same
exercise with a minimum allocation of 40\% in the equity asset class, the two
solutions are very different (see columns 6 and 8 in Table
\ref{tab:linear1-3}). The problem is that we assume that the constraint
$\sum_{i=1}^{n}x_{i}=1$ is managed by the set $\Omega$. This is not the case,
because the constraint $\sum_{i=1}^{n}x_{i}=1$ is managed by the Lagrange
multiplier $\lambda$ associated to the logarithmic barrier. Here, we face an
important issue called the scaling compatibility problem. This means that a
solution is acceptable if and only if the constraints $\Omega$ are
\textquotedblleft compatible\textquotedblright\ with the homogeneity property%
\footnote{For example, $\Omega=\left\{ x \in \mathbb{R}^{n} : x_{2} \geq 2
x_{1}\right\} $ is compatible with the scaling property.}.

\begin{table}[tbph]
\centering
\caption{Illustration of the scaling puzzle}
\label{tab:linear1-3}
\tableskip
\begin{tabular}{|c|cc:cc|cc:cc|}
\hline
$\phIII \sum_{i=5}^{8}x_{i}\geq 30\%$ & \multicolumn{2}{c:}{$\checkmark $} & & & & & & \\
$\phIII \sum_{i=1}^{4}x_{i}\leq 70\%$ & & & \multicolumn{2}{c|}{$\checkmark $} & & & & \\  \hdashline
$\phIII \sum_{i=5}^{8}x_{i}\geq 40\%$ & & & & & \multicolumn{2}{c:}{$\checkmark $} & & \\
$\phIII \sum_{i=1}^{4}x_{i}\leq 60\%$ & & & & & & & \multicolumn{2}{c|}{$\checkmark $} \\ \hline
Asset  & $x_i$ & $\mathcal{RC}_i^{\star}$ & $x_i$ & $\mathcal{RC}_i^{\star}$ & $x_i$ & $\mathcal{RC}_i^{\star}$ & $x_i$ & $\mathcal{RC}_i^{\star}$  \\
1 &      $25.78$ & ${\TsV}8.64$ &      $23.39$ & ${\TsV}6.50$ &      $24.09$ & ${\TsV}4.35$ &      $18.73$ & ${\TsV}2.01$  \\
2 &      $27.41$ & ${\TsV}8.64$ &      $24.34$ & ${\TsV}6.11$ &      $25.09$ & ${\TsV}4.35$ &      $19.08$ & ${\TsV}1.68$  \\
3 & ${\TsV}9.51$ & ${\TsV}8.64$ &      $12.46$ &      $10.98$ & ${\TsV}6.57$ & ${\TsV}4.35$ &      $12.48$ & ${\TsV}7.84$  \\
4 & ${\TsV}7.29$ & ${\TsV}8.64$ & ${\TsV}9.81$ &      $12.07$ & ${\TsV}4.24$ & ${\TsV}4.35$ & ${\TsV}9.71$ &      $10.43$  \\ \hdashline
5 & ${\TsV}7.06$ &      $15.91$ & ${\TsV}7.30$ &      $16.09$ & ${\TsV}8.74$ &      $18.59$ & ${\TsV}9.82$ &      $19.51$  \\
6 & ${\TsV}7.71$ &      $16.58$ & ${\TsV}7.66$ &      $16.09$ &      $10.75$ &      $21.87$ &      $10.27$ &      $19.51$  \\
7 & ${\TsV}9.23$ &      $18.14$ & ${\TsV}8.46$ &      $16.09$ &      $14.09$ &      $27.32$ &      $11.15$ &      $19.51$  \\
8 & ${\TsV}6.00$ &      $14.82$ & ${\TsV}6.57$ &      $16.09$ & ${\TsV}6.42$ &      $14.82$ & ${\TsV}8.76$ &      $19.51$  \\ \hline
\multicolumn{1}{|c|}{$\sigma\left(x\right)$ (in \%)} &
\multicolumn{2}{c:}{${\TsV}5.20$} & \multicolumn{2}{c|}{${\TsV}5.43$} & \multicolumn{2}{c:}{${\TsV}5.98$} & \multicolumn{2}{c|}{${\TsV}6.56$} \\
\multicolumn{1}{|c|}{$\mathcal{L}\left( x^{\star };\lambda ^{\star }\right) $ (in \%)} &
\multicolumn{2}{c:}{$13.29$} & \multicolumn{2}{c|}{$20.86$} & \multicolumn{2}{c:}{$10.68$} & \multicolumn{2}{c|}{$28.27$} \\
\multicolumn{1}{|c|}{Global minimum} &
\multicolumn{2}{c:}{$\checkmark$} & \multicolumn{2}{c|}{ } & \multicolumn{2}{c:}{$\checkmark$} & \multicolumn{2}{c|}{$ $} \\
\hline
\end{tabular}
\end{table}

How do we explain these results? The first reason is the choice of the starting
value for initializing the algorithm. It is obvious that the bisection
algorithm takes a road that depends on the initialization step. However, this
reason is not the primary answer, because we observe that we converge to the
same solution whatever the starting value or the algorithm when we consider box
or pointwise constraints. In fact, the main reason is the scaling property. In
this case, the convergence path is crucial, because we can obtain local
minima.\smallskip

When the constraints are incompatible with the homogeneity property of the risk
measure $\mathcal{R}\left( x\right) $, we may wonder if one solution is better
than the others. In order to answer this question, we recall that the numerical
algorithms solve the minimization problem $x^{\star }\left(
\lambda \right) =\arg \min \mathcal{L}\left( x;\lambda \right) $ where:%
\begin{equation*}
\mathcal{L}\left( x;\lambda \right) =\mathcal{R}\left( x\right) -\lambda
\sum_{i=1}^{n}b_{i}\ln x_{i}+\mathds{1}_{\Omega }\left( x\right)
\end{equation*}%
The idea is then to calculate $\mathcal{L}\left( x^{\star }\left( \lambda
^{\star }\right) ;\lambda ^{\star }\right) $ for the different solutions and to
take the solution that gives the lowest value. In Table \ref{tab:linear1-3},
the first and third portfolios are the best solutions. They have the lowest
volatility and Lagrange function.

\begin{remark}
Let us assume that the set of constraints includes the standard simplex:
$\mathcal{S}\subset\Omega$. By construction, the sum of weights is always equal
to $1$ whatever the value of the Lagrange multiplier $\lambda$:
\begin{equation*}
\sum_{i=1}^{n}x_{i}^{\star}\left(  \Omega,\lambda\right)  = 1
\end{equation*}
It gives the impression that there are two solutions. However, there is only
one solution which corresponds to the portfolio with the lowest risk measure.
\end{remark}

\section{Conclusion}

In this paper\footnote{This aim of this paper is also to help to popularize
large-scale optimization algorithms that are very popular in machine learning,
but are not well-known in finance. However, we think that they are relevant for
many financial applications, in particular for portfolio optimization. Our
paper which is focused on risk-budgeting optimization can then be seen as a
companion work of Bourgeron \textsl{et al.} (2018) which focuses on
mean-variance optimization.}, we propose an approach to find the risk budgeting
portfolio when we impose some constraints. The underlying idea is to consider
the logarithmic barrier problem and to use recent optimization algorithms in
order to find the numerical solution. In particular, we use cyclical coordinate
descent (CCD), alternative direction method of multipliers (ADMM), proximal
operators and Dykstra's algorithm.\smallskip

We provide different examples that are focused on the ERC portfolio. This
portfolio is very interesting since it imposes the same risk contribution
between the assets of the investment universe. We may then wonder what does an
ERC portfolio mean when we impose some constraints. Most of the times, we
observe that the ERC property continues to be satisfied for the assets that are
not impacted by the constraints. This type of approach is very appealing when
we would like to manage the liquidity of a portfolio, the small cap bias of
risk-based indices or the turnover of a risk parity fund.\smallskip

This study also highlights the importance of the homogeneity property of RB
portfolios. Roncalli (2015) has already pointed out that the risk measure must
be coherent, which can be incompatible when imposing constraints. The
homogeneity property really asks the question of the compatibility between risk
budgeting allocation and imposing some constraints. However, even if the
calibration of the Lagrange multiplier $\lambda$ remains an issue in some
cases, our approach extends the optimization framework defined by Richard and
Roncalli (2015), and reinforces that idea that risk budgeting and risk
minimization are highly connected.

\clearpage

\appendix

\section*{Appendix}

\section{Optimization algorithms}

\subsection{ADMM algorithm}
\label{appendix:section-admm}

The alternating direction method of multipliers (ADMM) is an algorithm
introduced by Gabay and Mercier (1976) to solve problems which can be expressed
as\footnote{We follow the standard presentation of Boyd \textsl{et al.} (2011)
on ADMM.}:
\begin{eqnarray}
\left\{ x^{\star },z^{\star }\right\}  &=&\arg \min f\left( x\right)
+g\left( z\right)   \label{eq:appendix-admm1} \\
&\text{s.t.}&Ax+Bz-c=0  \notag
\end{eqnarray}%
where $A\in \mathbb{R}^{p\times n}$, $B\in \mathbb{R}^{p\times m}$, $c\in
\mathbb{R}^{p}$, and the functions $f:\mathbb{R}^{n}\rightarrow \mathbb{R}%
\cup \{+\infty \}$ and $g:\mathbb{R}^{m}\rightarrow \mathbb{R}\cup \{+\infty
\}$ are proper closed convex functions. Boyd \textsl{et al.} (2011) show that
the ADMM algorithm consists of three steps:

\begin{enumerate}
\item The $x$-update is:
\begin{equation}
x^{\left( k\right) }=\arg \min \left\{ f\left( x\right) +\frac{\varphi }{2}%
\left\Vert Ax+Bz^{\left( k-1\right) }-c+u^{\left( k-1\right) }\right\Vert
_{2}^{2}\right\}   \label{eq:appendix-admm2a}
\end{equation}

\item The $z$-update is:%
\begin{equation}
z^{\left( k\right) }=\arg \min \left\{ g\left( z\right) +\frac{\varphi }{2}%
\left\Vert Ax^{\left( k\right) }+Bz-c+u^{\left( k-1\right) }\right\Vert
_{2}^{2}\right\}   \label{eq:appendix-admm2b}
\end{equation}

\item The $u$-update is:%
\begin{equation}
u^{\left( k\right) }=u^{\left( k-1\right) }+\left( Ax^{\left( k\right)
}+Bz^{\left( k\right) }-c\right)   \label{eq:appendix-admm2c}
\end{equation}
\end{enumerate}
In this approach, $u^{\left( k\right) }$ is the dual variable of the primal
residual $r=Ax+Bz-c$ and $\varphi $ is the $\boldsymbol{\ell}_{2}$ penalty variable. In the
paper, we use the notations $f^{\left( k\right) }\left( x\right) $ and
$g^{\left( k\right) }\left( z\right) $ when referring to the objective
functions that are defined in the $x$- and $z$-steps.

\subsection{Proximal operator}
\label{appendix:section-proximal}

In what follows, we give the main results that are summarized in Bourgeron
\textsl{et al.} (2017). Let $f:\mathbb{R}^{n}\rightarrow \mathbb{R}\cup \left\{
+\infty \right\} $ be a proper closed convex function. The proximal operator
$\mathbf{prox}_{f}\left( v\right) :\mathbb{R}^{n}\rightarrow
\mathbb{R}^{n}$ is defined by:%
\begin{equation}
\mathbf{prox}_{f}\left( v\right) =x^{\star }=\arg \min\nolimits_{x}\left\{
f\left( x\right) +\frac{1}{2}\left\Vert x-v\right\Vert _{2}^{2}\right\}
\label{eq:appendix-prox1}
\end{equation}%
Since the function $f_{v}\left( x\right) =f\left( x\right) +\dfrac{1}{2}%
\left\Vert x-v\right\Vert _{2}^{2}$ is strongly convex, it has a unique minimum
for every $v\in \mathbb{R}^{n}$ (Parikh and Boyd, 2014). For example, if we
consider the logarithmic barrier function $f\left( x\right) =-\ln x$, we have:
\begin{eqnarray*}
f\left( x\right) +\frac{1}{2}\left\Vert x-v\right\Vert _{2}^{2} &=&-\ln x+%
\frac{1}{2}\left( x-v\right) ^{2} \\
&=&-\ln x+x^{2}-xv+\frac{1}{2}v^{2}
\end{eqnarray*}
The first-order condition is $-x^{-1}+2x-v=0$. We obtain two roots with
opposite signs. Since the logarithmic function is defined for $x>0$, we deduce
that the proximal operator is:
\begin{equation*}
\mathbf{prox}_{f}(v)=\frac{v+\sqrt{v^{2}+4}}{2}
\end{equation*}%
More generally, if we consider $f\left( x\right) =-\lambda \sum_{i=1}^{n}\ln
x_{i}$, we have:
\begin{equation*}
\left( \mathbf{prox}_{f}(v)\right) _{i}=\frac{v_{i}+\sqrt{v_{i}^{2}+4\lambda}}{2}
\end{equation*}
\smallskip

Let us now consider some special cases. If we assume that $f\left( x\right)
=\mathds{1}_{\Omega }\left( x\right) $
where $\Omega $ is a convex set, we have:%
\begin{eqnarray}
\mathbf{prox}_{f}\left( v\right)  &=&\arg \min\nolimits_{x}\left\{ \mathds{1}%
_{\Omega }\left( x\right) +\frac{1}{2}\left\Vert x-v\right\Vert
_{2}^{2}\right\}   \notag \\
&=&\mathcal{P}_{\Omega }\left( v\right)   \label{eq:appendix-prox2}
\end{eqnarray}%
where $\mathcal{P}_{\Omega }\left( v\right) $ is the standard projection. Here,
we give the results of Parikh and Boyd (2014) for some simple polyhedra:%
\begin{equation*}
\begin{tabular}{cc}
\hline
$\Omega $                         & $\mathcal{P}_{\Omega }\left( v\right) $                                          \\ \hline
$Ax=B$                            & $v-A^{\dagger }\left( Av-B\right) $                                              \\
$a^{\top }x=b$                    & $v-\dfrac{\left( a^{\top }v-b\right) }{\left\Vert a\right\Vert _{2}^{2}}a$       \\
$c^{\top }x\leqslant d$           & $v-\dfrac{\left( c^{\top }v-d\right) _{+}}{\left\Vert c\right\Vert _{2}^{2}}c$   \\
$x^{-}\leqslant x\leqslant x^{+}$ & $\mathcal{T}\left( v;x^{-},x^{+}\right) $                                        \\ \hline
\end{tabular}%
\end{equation*}%
where $A^{\dagger }$ is the Moore-Penrose pseudo-inverse of $A$, and $\mathcal{T}\left( v;x^{-},x^{+}\right) $ is the truncation operator:%
\begin{eqnarray*}
\mathcal{T}\left( v;x^{-},x^{+}\right)  &=&v\odot \mathds{1}\left\{
x^{-}\leqslant v\leqslant x^{+}\right\} + \\
&&x^{-}\odot \mathds{1}\left\{ v<x^{-}\right\} + \\
&&x^{+}\odot \mathds{1}\left\{ v>x^{+}\right\}
\end{eqnarray*}
In the case of complex polyhedra, the reader can find analytical formulas and
numerical algorithms in Parikh and Boyd (2014), and Combettes and Pesquet
(2011).\smallskip

We also have:%
\begin{eqnarray*}
x^{\star } &=&\arg \min \frac{1}{2}\left\Vert x-v\right\Vert _{2}^{2} \\
\text{} &\text{s.t.}&x\in \Omega
\end{eqnarray*}%
If we define $\Omega $ as follows:%
\begin{equation*}
\Omega =\left\{ x\in \mathbb{R}^{n}:Ax=B,Cx\leq D,x^{-}\leq x\leq
x^{+}\right\}
\end{equation*}%
we obtain:%
\begin{eqnarray*}
x^{\star } &=&\arg \min \frac{1}{2}x^{\top }x-v^{\top }x \\
\text{{}} &\text{s.t.}&\left\{
\begin{array}{l}
Ax=B \\
Cx\leq D \\
x^{-}\leq x\leq x^{+}%
\end{array}%
\right.
\end{eqnarray*}%
Imposing linear constraints is then equivalent to solving a standard QP problem.
\smallskip

We now consider the case of norm functions. For that, we need a preliminary
result. In the case of the pointwise maximum function $f\left( x\right)
=\max x$, we have:%
\begin{equation}
\mathbf{prox}_{\lambda f}\left( v\right) =\min \left( v,s^{\star }\right)
\label{eq:proximal-max}
\end{equation}%
where $s^{\star }$ is the solution of the following equation:%
\begin{equation*}
s^{\star }=\left\{ s\in \mathbb{R}:\sum_{i=1}^{n}\left( v_{i}-s\right)
_{+}=\lambda \right\}
\end{equation*}%
If we assume that $f\left( x\right) =\left\Vert x\right\Vert _{p}$, we obtain%
\footnote{%
The proximal operator $S_{\lambda }\left( v\right) $ is known as the soft
thresholding operator. In particular, it is used for solving lasso
regression problems (Friedman \textsl{et al.}, 2010).}:
\begin{equation*}
\begin{tabular}{cc}
\hline
$p$ & $\mathbf{prox}_{\lambda f}\left( v\right) $ \\ \hline
$p=1$ & $S_{\lambda }\left( v\right) =\left( \left\vert v\right\vert
-\lambda \mathbf{1}\right) _{+}\odot \limfunc{sign}\left( v\right) $ \\
$p=2$ & $\left( 1-\dfrac{1}{\max \left( \lambda ,\left\Vert v\right\Vert
_{2}\right) }\right) v$ \\
$p=\infty $ & $\mathbf{prox}_{\lambda \max }\left( \left\vert v\right\vert
\right) \odot \limfunc{sign}\left( v\right) $ \\ \hline
\end{tabular}%
\end{equation*}%
An important property of the proximal operator is the Moreau decomposition
theorem:%
\begin{equation*}
\mathbf{prox}_{f}\left( v\right) +\mathbf{prox}_{f^{\ast }}\left( v\right) =v
\end{equation*}%
where $f^{\ast }$ is the convex conjugate of $f$. If $f\left( x\right) $ is
a $\boldsymbol{\ell }_{p}$-norm function, then $f^{\ast }\left( x\right) =%
\mathds{1}_{\mathcal{B}_{p}}\left( x\right) $ where $\mathcal{B}_{p}$ is the
$\boldsymbol{\ell }_{p}$ unit ball. Since we have $\mathbf{prox}_{f^{\ast
}}\left( v\right) =\mathcal{P}_{\mathcal{B}_{p}}\left( v\right) $, we deduce
that:%
\begin{equation*}
\mathbf{prox}_{f}\left( v\right) +\mathcal{P}_{\mathcal{B}_{p}}\left(
v\right) =v
\end{equation*}%
More generally, we have:%
\begin{equation*}
\mathbf{prox}_{\lambda f}\left( v\right) +\lambda \mathcal{P}_{\mathcal{B}%
_{p}}\left( \frac{v}{\lambda }\right) =v
\end{equation*}%
It follows that the projection on $\boldsymbol{\ell }_{p}$ ball can be
deduced from the proximal operator of the $\boldsymbol{\ell }_{p}$-norm function. Let
$\mathcal{B}_{p}\left( c,\lambda \right) =\left\{ x\in \mathbb{R}%
^{n}:\left\Vert x-c\right\Vert _{p}\leq \lambda \right\} $ be the $%
\boldsymbol{\ell }_{p}$ ball with center $c$ and radius $\lambda $. We
obtain:
\begin{equation*}
\begin{tabular}{cc}
\hline
$p$ & $\mathcal{P}_{\mathcal{B}_{p}\left( \mathbf{0},\lambda \right) }\left(
v\right) $ \\ \hline
$p=1$ & $v-\mathbf{prox}_{\lambda \max }\left( \left\vert v\right\vert
\right) \odot \limfunc{sign}\left( v\right) $ \\
$p=2$ & $v-\mathbf{prox}_{\lambda \left\Vert {\cdot}\right\Vert _{2}}\left(
\left\vert v\right\vert \right) $ \\
$p=\infty $ & $\mathcal{T}\left( v;-\lambda ,\lambda \right) $ \\ \hline
\end{tabular}%
\end{equation*}%
In the case where the center $c$ is not equal to $\mathbf{0}$, we consider
the translation property:%
\begin{equation*}
\mathbf{prox}_{g}\left( v\right) =\mathbf{prox}_{f}\left( v+c\right) -c
\end{equation*}%
where $g\left( x\right) =f\left( x+c\right) $. Since we have the equivalence
$\mathcal{B}_{p}\left( \mathbf{0},\lambda \right) =\left\{ x\in \mathbb{R}%
^{n}:f\left( x\right) \leq \lambda \right\} $ where $f\left( x\right)
=\left\Vert x\right\Vert _{p}$, we deduce that:%
\begin{equation*}
\mathcal{P}_{\mathcal{B}_{p}\left( c,\lambda \right) }\left( v\right) =%
\mathcal{P}_{\mathcal{B}_{p}\left( \mathbf{0},\lambda \right) }\left(
v-c\right) +c
\end{equation*}

\subsection{Dykstra's algorithm}
\label{appendix:section-dykstra}

We consider the following proximal problem:%
\begin{equation*}
x^{\star }=\mathbf{prox}_{f}\left( v\right)
\end{equation*}%
where $f\left( x\right) =\mathds{1}_{\Omega }\left( x\right) $ and:%
\begin{equation*}
\Omega =\Omega _{1} \cap \Omega _{2} \cap \cdots \cap \Omega _{m}
\end{equation*}%
The solution can be found thanks to Dykstra's algorithm (Dykstra, 1983;
Bauschke and Borwein, 1994), which consists in the following two steps until
convergence:
\begin{enumerate}
\item The $x$-update is:%
\begin{equation*}
x^{\left( k\right) }=\mathcal{P}_{\Omega _{\limfunc{mod}\left( k,m\right)
}}\left( x^{\left( k-1\right) }+z^{\left( k-m\right) }\right)
\end{equation*}
\item The $z$-update is:%
\begin{equation*}
z^{\left( k\right) }=x^{\left( k-1\right) }+z^{\left( k-m\right) }-x^{\left(
k\right) }
\end{equation*}
\end{enumerate}
where $x^{\left( 0\right) }=v$, $z^{\left( k\right) }=\mathbf{0}$ for $k<0$ and
$\limfunc{mod}\left( k,m\right) $ denotes the modulo operator taking values in
$\left\{ 1,\ldots ,m\right\} $.\smallskip

Let us consider the case $\Omega =\left\{ x\in \mathbb{R}^{n}:Cx\leq D\right\}
$ where the number of inequality constraints is equal to $m$. We
can write:%
\begin{equation*}
\Omega =\Omega _{1}\cap \Omega _{2}\cap \cdots \cap \Omega _{m}
\end{equation*}%
where $\Omega _{j}=\left\{ x\in \mathbb{R}^{n}:c_{\left( j\right) }^{\top
}x\leq d_{\left( j\right) }\right\} $, $c_{\left( j\right) }^{\top }$
corresponds to the $j^{\mathrm{th}}$ row of $C$ and $d_{\left( j\right) }$ is
the $j^{\mathrm{th}}$ element of $D$. We follow Tibshirani (2017) to define the
corresponding algorithm. In particular, we introduce two iteration indices $j$
and $k$. The index $j$ refers to the constraint number, whereas the index $k$
refers to the main loop. Algorithm \ref{alg:dykstra1} describes the Dykstra's
approach for solving this proximal problem.\smallskip

\begin{algorithm}[tbh]
\caption{Dykstra's algorithm for solving the proximal problem with inequality
constraints} \label{alg:dykstra1}
\begin{algorithmic}
\STATE  The goal is to compute the solution $x^{\star }=\mathbf{prox}_{f}\left(
v\right) $ where $f\left( x\right) =\mathds{1}_{\Omega }\left( x\right) $ and $\Omega =\left\{ x\in \mathbb{R}^{n}:Cx\leq D\right\} $
\STATE  We initialize $x_{m}^{\left( 0\right) }\leftarrow v$
\STATE  We set $z_{1}^{\left( 0\right) }\leftarrow \mathbf{0}, \ldots, z_{m}^{\left( 0\right)
}\leftarrow \mathbf{0}$
\STATE  We note $k_{\max }$ the maximum number of iterations
\FOR {$k = 1 : k_{\max}$}
    \STATE $x_{0}^{\left( k\right) }\leftarrow x_{m}^{\left( k-1\right) }$
    \FOR {$j=1:m$}
        \STATE  The $x$-update is:%
            \begin{eqnarray*}
                x_{j}^{\left( k\right) } &=&\mathcal{P}_{\Omega _{j}}\left( x_{j-1}^{\left(
                k\right) }+z_{j}^{\left( k-1\right) }\right)  \\
                &=&x_{j-1}^{\left( k\right) }+z_{j}^{\left( k-1\right) }-\frac{\left(
                c_{\left( j\right) }^{\top }x_{j-1}^{\left( k\right) }+c_{\left( j\right)
                }^{\top }z_{j}^{\left( k-1\right) }-d_{\left( j\right) }\right) _{+}}{%
                \left\Vert c_{\left( j\right) }\right\Vert _{2}^{2}}c_{\left( j\right) }
            \end{eqnarray*}
        \STATE The $z$-update is:%
            \begin{equation*}
                z_{j}^{\left( k\right) }=x_{j-1}^{\left( k\right) }+z_{j}^{\left(
                k-1\right) }-x_{j}^{\left( k\right) }
            \end{equation*}
    \ENDFOR
    \IF{$x_{m}^{\left( k\right) }=x_{0}^{\left( k-1\right) }$}
        \STATE Break
    \ENDIF
\ENDFOR
\RETURN $x^{\star }\leftarrow x_{m}^{\left( k\right) }$
\end{algorithmic}
\end{algorithm}

If we define $\Omega $ as follows:%
\begin{equation*}
\Omega =\left\{ x\in \mathbb{R}^{n}:Ax=B,Cx\leq D,x^{-}\leq x\leq
x^{+}\right\}
\end{equation*}%
we decompose $\Omega $ as the intersection of three basic convex sets:%
\begin{equation*}
\Omega =\Omega _{1}\cap \Omega _{2}\cap \Omega _{3}
\end{equation*}%
where $\Omega _{1}=\left\{ x\in \mathbb{R}^{n}:Ax=B\right\} $, $\Omega
_{2}=\left\{ x\in \mathbb{R}^{n}:Cx\leq D\right\} $ and $\Omega _{3}=\left\{
x\in \mathbb{R}^{n}:x^{-}\leq x\leq x^{+}\right\} $. Using Dykstra's algorithm
is equivalent to formulating Algorithm \ref{alg:dykstra2}.

\begin{remark}
An alternative approach is to write the constraints in the following way%
\footnote{We use the following properties:
\begin{equation*}
Ax=B\Leftrightarrow Ax\leq B\text{ and }Ax\geq B
\end{equation*}%
and:%
\begin{equation*}
x^{-}\leq x\leq x^{+}\Leftrightarrow -x\leq -x^{-}\text{ and }x\leq x^{+}
\end{equation*}}:
\begin{equation*}
\Omega =\left\{ x\in \mathbb{R}^{n}:C^{\star }x\leq D^{\star }\right\}
\end{equation*}%
where $C^{\star }=\left( A,-A,C,-I_{n},I_{n}\right) $ and $D^{\star }=\left(
B,-B,D,-x^{-},x^{+}\right) $. Therefore, we can use Algorithm \ref{alg:dykstra1} to find the
solution.
\end{remark}

\subsection{Proximal operator of the risk budgeting logarithmic barrier}
\label{appendix:section-log-barrier}

We have:%
\begin{equation*}
z^{\left( k\right) }=\arg \min g^{\left( k\right) }\left( z\right)
\end{equation*}%
where:%
\begin{eqnarray*}
g^{\left( k\right) }\left( z\right)  &=&g\left( z\right) +\frac{\varphi }{2}%
\left\Vert x^{\left( k\right) }-z+u^{\left( k-1\right) }\right\Vert _{2}^{2}
\\
&=&-\lambda \sum_{i=1}^{n}b_{i}\ln z_{i}+\frac{\varphi }{2}%
\sum_{i=1}^{n}\left( x_{i}^{\left( k\right) }-z_{i}+u_{i}^{\left( k-1\right)
}\right) ^{2} \\
&=&\sum_{i=1}^{n}\left( \frac{\varphi }{2}\left( x_{i}^{\left( k\right)
}-z_{i}+u_{i}^{\left( k-1\right) }\right) ^{2}-\lambda b_{i}\ln z_{i}\right)
\end{eqnarray*}%
The first-order condition is:%
\begin{equation*}
\frac{\partial \,g^{\left( k\right) }\left( z\right) }{\partial \,z_{i}}%
=-\varphi \left( x_{i}^{\left( k\right) }-z_{i}+u_{i}^{\left( k-1\right)
}\right) -\lambda b_{i}\frac{1}{z_{i}}=0
\end{equation*}%
We deduce that $z_{i}^{\left( k\right) }$ is the solution of the quadratic
equation:%
\begin{equation*}
\left\{
\begin{array}{l}
\varphi z_{i}^{2}-\varphi \left( x_{i}^{\left( k\right) }+u_{i}^{\left(
k-1\right) }\right) z_{i}-\lambda b_{i}=0 \\
z_{i}>0%
\end{array}%
\right.
\end{equation*}%
{\begin{algorithm}[t]
\caption{Dykstra's algorithm for solving the proximal problem with general linear constraints}
\label{alg:dykstra2}
\begin{algorithmic}
\STATE  The goal is to compute the solution $x^{\star }=\mathbf{prox}%
_{f}\left( v\right) $ where $f\left( x\right) =\mathds{1}_{\Omega }\left(
x\right) $ and $\Omega =\left\{ x\in \mathbb{R}^{n}:Ax=B,Cx\leq D,x^{-}\leq
x\leq x^{+}\right\} $
\STATE  We initialize $x_{m}^{\left( 0\right) }\leftarrow v$
\STATE  We set $z_{1}^{\left( 0\right) }\leftarrow \mathbf{0}$, $z_{2}^{\left( 0\right) }\leftarrow \mathbf{0}$ and
        $z_{3}^{\left( 0\right) }\leftarrow \mathbf{0}$
\STATE  We note $k_{\max }$ the maximum number of iterations
\FOR {$k = 1 : k_{\max}$}
    \STATE $x_{0}^{\left( k\right) }\leftarrow x_{m}^{\left( k-1\right) }$
    \STATE For the set $\Omega _{1}$, we have:%
        \begin{equation*}
        \left\{
        \begin{array}{l}
        x_{1}^{\left( k\right) }\leftarrow x_{0}^{\left( k\right) }+z_{1}^{\left(
        k-1\right) }-A^{\dag }\left( Ax_{0}^{\left( k\right) }+Az_{1}^{\left(
        k-1\right) }-B\right)  \\
        z_{1}^{\left( k\right) }\leftarrow x_{0}^{\left( k\right) }+z_{1}^{\left(
        k-1\right) }-x_{1}^{\left( k\right) }%
        \end{array}%
        \right.
        \end{equation*}
    \STATE For the set $\Omega _{2}$, we have\footnotemark :
        \begin{equation*}
        \left\{
        \begin{array}{l}
        x_{2}^{\left( k\right) }\leftarrow \mathcal{P}_{\Omega _{2}}\left(
        x_{1}^{\left( k\right) }+z_{2}^{\left( k-1\right) }\right)  \\
        z_{2}^{\left( k\right) }\leftarrow x_{1}^{\left( k\right) }+z_{2}^{\left(
        k-1\right) }-x_{2}^{\left( k\right) }%
        \end{array}%
        \right.
        \end{equation*}%
    \STATE For the set $\Omega _{3}$, we have:%
        \begin{equation*}
        \left\{
        \begin{array}{l}
        x_{3}^{\left( k\right) }\leftarrow \mathcal{T}\left( x_{2}^{\left( k\right)
        }+z_{3}^{\left( k-1\right) };x^{-},x^{+}\right)  \\
        z_{3}^{\left( k\right) }\leftarrow x_{2}^{\left( k\right) }+z_{3}^{\left(
        k-1\right) }-x_{3}^{\left( k\right) }%
        \end{array}%
        \right.
        \end{equation*}
    \IF{$x_{3}^{\left( k\right) }=x_{0}^{\left( k\right) }$}
        \STATE Break
    \ENDIF
\ENDFOR
\RETURN $x^{\star }\leftarrow x_{3}^{\left( k\right) }$
\end{algorithmic}
\end{algorithm}
\footnotetext{This step is done using Algorithm \ref{alg:dykstra1}.}}
We have:%
\begin{equation*}
\Delta =\varphi ^{2}\left( x_{i}^{\left( k\right) }+u_{i}^{\left( k-1\right)
}\right) ^{2}+4\varphi \lambda b_i
\end{equation*}%
Since $\Delta >0$ and $-\lambda \varphi b_{i}<0$, we have two roots with
opposite signs. Therefore, the solution is equal to:%
\begin{equation*}
z_{i}^{\left( k\right) }=\frac{\varphi \left( x_{i}^{\left( k\right)
}+u_{i}^{\left( k-1\right) }\right) +\sqrt{\varphi ^{2}\left( x_{i}^{\left(
k\right) }+u_{i}^{\left( k-1\right) }\right) ^{2}+4\varphi \lambda b_i}}{%
2\varphi }
\end{equation*}

\subsection{CCD algorithm with separable constraints}
\label{appendix:section-wright}

The coordinate update proposed by Nesterov (2012) and Wright (2015) is:
\begin{equation*}
x_{i}^{\star }=\arg \min \left( x-x_{i}\right) g_{i}+\frac{1}{2\eta }\left(
x-x_{i}\right) ^{2}+\xi \cdot \mathds{1}_{\Omega _{i}}\left( x_{i}\right)
\end{equation*}%
where $\xi $ is a positive scalar, $\eta >0$ is the stepsize of the
gradient descent and $g_i$ is the first-derivative of the function with respect to $x_i$:%
\begin{equation*}
g_{i}=-\pi _{i}+c\frac{\left( \Sigma x\right) _{i}}{\sqrt{x^{\top }\Sigma x}}%
-\lambda \frac{b_{i}}{x_{i}}
\end{equation*}%
The objective function is equivalent to:%
\begin{eqnarray*}
(\ast ) &=&\left( x-x_{i}\right) g_{i}+\frac{1}{2\eta }\left( x-x_{i}\right)
^{2}+\xi \cdot \mathds{1}_{\Omega _{i}}\left( x_{i}\right)  \\
&=&\frac{1}{2\eta }\left( \left( x-x_{i}\right) ^{2}+2\left( x-x_{i}\right)
\eta g_{i}\right) +\xi \cdot \mathds{1}_{\Omega _{i}}\left( x_{i}\right)  \\
&=&\frac{1}{2\eta }\left( x-x_{i}+\eta g_{i}\right) ^{2}+\xi \cdot \mathds{1}%
_{\Omega _{i}}\left( x_{i}\right) -\frac{\eta }{2}g_{i}^{2}
\end{eqnarray*}%
By taking $\xi =\eta ^{-1}$, we deduce that:%
\begin{eqnarray*}
x_{i}^{\star } &=&\arg \min \mathds{1}_{\Omega _{i}}\left( x_{i}\right) +%
\frac{1}{2}\left\Vert x-\left( x_{i}-\eta g_{i}\right) \right\Vert ^{2} \\
&=&\mathbf{prox}_{\psi }\left( x_{i}-\eta g_{i}\right)
\end{eqnarray*}%
where $\psi \left( x\right) =\mathds{1}_{\Omega _{i}}\left( x\right) $.

\subsection{Accelerated bisection algorithm}
\label{appendix:section-bisection}

The classic bisection algorithm consists in calculating the solution $%
x^{\star }\left( \lambda \right) =\arg \min \mathcal{L}\left( x;\lambda \right)
$ and updating the bounds of the interval $\left[ a_{\lambda },b_{\lambda
}\right] $ that contains the solution $\lambda ^{\star }$ such that
$\sum\limits_{i=1}^{n}x^{\star }\left( \lambda \right) =1$. One of the issues
is that $x^{\star }\left( \lambda \right) $ is obtained by an optimization
algorithm and is not an analytical formula. This means that we
can write:%
\begin{equation*}
x^{\star }\left( \lambda ;x^{\left( 0\right) }\right) =\arg \min \mathcal{L}%
\left( x;\lambda ,x^{\left( 0\right) }\right)
\end{equation*}%
where $x^{\left( 0\right) }$ is the starting value of the ADMM or CCD
algorithm. Therefore, we can update the starting value $x^{\left( 0\right) }$
of the algorithm at each iteration of the bisection method. This helps to
accelerate the convergence of the $x$-update. In practice, we can replace
Algorithm \ref{alg:algorithm1} by Algorithm \ref{alg:algorithm5}.

\begin{algorithm}[tbph]
\begin{algorithmic}
\STATE The goal is to compute the optimal Lagrange multiplier $\lambda ^{\star }$ and the solution $x^{\star }\left( \mathcal{S},\Omega \right) $
\STATE We consider two scalars $a_{\lambda }$ and $b_{\lambda }$ such that $a_{\lambda }<b_{\lambda }$ and $\lambda^{\star }\in \left[ a_{\lambda },b_{\lambda }\right] $
\STATE We note $\varepsilon _{\lambda }$ the convergence criterion of the bisection algorithm (e.g. $10^{-8}$)
\STATE We note $x^{\left( 0\right) }$ the starting value of the ADMM/CCD algorithm
\REPEAT
    \STATE We calculate $\lambda =\dfrac{a_{\lambda }+b_{\lambda }}{2}$
    \STATE We compute $x^{\star }\left( \lambda ;x^{\left( 0\right) }\right) $ the solution of the minimization problem:
            \begin{equation*}
                x^{\star }\left( \lambda ;x^{\left( 0\right) }\right) =\arg \min \mathcal{L}\left( x;\lambda ,x^{\left( 0\right) }\right)
            \end{equation*}
    \IF{$\sum_{i=1}^{n}x_{i}^{\star }\left( \lambda ;x^{\left( 0\right) }\right) <1$}
    \STATE $a_{\lambda}\leftarrow \lambda $
    \ELSE
    \STATE $b_{\lambda }\leftarrow \lambda $
    \ENDIF
    \STATE $x^{\left( 0\right) }\leftarrow x^{\star }\left( \lambda ;x^{\left( 0\right)}\right) $
\UNTIL{$\left\vert \sum\limits_{i=1}^{n}x_{i}^{\star }\left( \lambda ;x^{\left(0\right) }\right) -1\right\vert \leq \varepsilon _{\lambda }$}
\RETURN $\lambda ^{\star }\leftarrow \lambda $ and $x^{\star }\left( \mathcal{S},\Omega \right) \leftarrow x^{\star }\left(\lambda ^{\star };x^{\left( 0\right) }\right) $
\end{algorithmic}
\caption{Accelerated bisection method}
\label{alg:algorithm5}
\end{algorithm}

\subsection{Adaptive penalization parameter}
\label{appendix:section-adaptive}

The convergence of the ADMM algorithm holds regardless of the value of the
penalization parameter $\varphi >0$. But the choice of $\varphi $ affects the
speed of convergence. In practice, the penalization parameter $\varphi $ may be
changed at each iteration, implying that $\varphi $ is replaced by $\varphi
^{\left( k\right) }$. This may improve the convergence and make the performance
of the ADMM algorithm less dependent of the initial choice $\varphi ^{\left(
0\right) } $. To update $\varphi ^{\left( k\right) }$ in practice, He
\textsl{et al.} (2000) and Wang and Liao (2001) provide a simple and efficient
scheme. Let $r^{\left( k\right) }=Ax^{\left( k\right) }+Bz^{\left( k\right)
}-c$ and $s^{\left( k\right) }=\varphi A^{\top }B\left( z^{\left( k\right)
}-z^{\left( k-1\right) }\right) $ be the primal and dual residual variables
(Boyd \textsl{et al.}, 2011). On the one hand, the $x$ and $z$-updates
essentially comes from placing a penalty on $\left\Vert r^{\left( k\right)
}\right\Vert _{2}^{2}$. As a consequence, if $\varphi ^{\left( k\right) }$ is
large, $\left\Vert r^{\left( k\right) }\right\Vert _{2}^{2}$ tends to be small.
On the other hand, $s^{\left( k\right) }$ depends linearly on $\varphi $. As a
consequence, if $\varphi ^{\left( k\right) }$ is small, $\left\Vert s^{\left(
k\right) }\right\Vert _{2}^{2}$ is small and $\left\Vert r^{\left( k\right)
}\right\Vert _{2}^{2}$ may be large. To keep $\left\Vert r^{\left( k\right)
}\right\Vert _{2}^{2}$ and $\left\Vert s^{\left( k\right) }\right\Vert
_{2}^{2}$ within a factor $\mu $, one may consider the following scheme:

\begin{enumerate}
\item If $\left\Vert r^{\left( k\right) }\right\Vert _{2}^{2}>\mu \left\Vert
s^{\left( k\right) }\right\Vert _{2}^{2}$, the $\varphi $-update is:%
\begin{equation*}
\left\{
\begin{array}{l}
\varphi ^{\left( k+1\right) }\leftarrow \tau \varphi ^{\left( k\right) } \\
u^{\left( k+1\right) }\leftarrow \dfrac{u^{\left( k+1\right) }}{\tau }%
\end{array}%
\right.
\end{equation*}

\item If $\left\Vert s^{\left( k\right) }\right\Vert _{2}^{2}>\mu \left\Vert
r^{\left( k\right) }\right\Vert _{2}^{2}$, we have:%
\begin{equation*}
\left\{
\begin{array}{l}
\varphi ^{\left( k+1\right) }\leftarrow \dfrac{\varphi ^{\left( k\right) }}{%
\tau ^{\prime }} \\
u^{\left( k+1\right) }\leftarrow \tau ^{\prime }u^{\left( k+1\right) }%
\end{array}%
\right.
\end{equation*}

\item Otherwise, the penalization parameter remains the same $\varphi
^{\left( k+1\right) }\leftarrow \varphi ^{\left( k\right) }$, implying that
we do not rescale the dual variable $u^{\left( k+1\right) }$.
\end{enumerate}
The previous scheme corresponds to the $\varphi $-update and must be placed
after the $u$-update of the ADMM algorithm. In practice, we use the following
default values: $\varphi ^{\left( 0\right) }=1$, $u^{\left( 0\right) }=0$, $\mu
=10^{6}$ and $\tau =\tau ^{\prime }=2$.

\section{Implementation}

A Python implementation is available on the following webpage:
\begin{center}
\url{https://github.com/jcrichard}
\end{center}

\end{document}